\documentclass[prx,letterpaper,aps,10pt,superscriptaddress,twocolumn,floatfix,showpacs]{revtex4-1}
\usepackage{graphicx}
\usepackage{amsmath}
\usepackage{amsfonts}
\usepackage{float}
\usepackage{amssymb}
\usepackage{epsfig}
\usepackage{epstopdf}
\DeclareGraphicsExtensions{.pdf,.eps,.png,.jpg,.mps}
\usepackage[pdftex]{color}
\usepackage{amsmath,graphicx,amssymb,braket,xcolor,subfigure,upgreek}
\usepackage[colorlinks, linkcolor=blue, citecolor=blue, urlcolor=blue, breaklinks=true]{hyperref}
\usepackage{microtype}
\usepackage{bbm}
\usepackage{color}
\usepackage{physics}
\usepackage{comment}
\usepackage{dsfont}
\usepackage{soul}

\begin{document}

\title{The Ising model in a light-induced quantized transverse field}
\author{Jonas Rohn}
\affiliation{Department of Physics, FAU Erlangen-Nuremberg, Staudtstra{\ss}e 7,
D-91058 Erlangen, Germany}
\author{Max H\"ormann}
\affiliation{Department of Physics, FAU Erlangen-Nuremberg, Staudtstra{\ss}e 7,
D-91058 Erlangen, Germany}
\author{Claudiu Genes}
\affiliation{Max Planck Institute for the Science of Light, Staudtstra{\ss}e 2,
D-91058 Erlangen, Germany}
\affiliation{Department of Physics, FAU Erlangen-Nuremberg, Staudtstra{\ss}e 7,
D-91058 Erlangen, Germany}
\author{Kai Phillip Schmidt}
\affiliation{Department of Physics, FAU Erlangen-Nuremberg, Staudtstra{\ss}e 7,
D-91058 Erlangen, Germany}
\date{\today}

\begin{abstract}
We investigate the influence of light-matter interactions on correlated quantum matter by studying the paradigmatic Ising model subject to a quantum Rabi coupling. This type of coupling to a confined, spatially delocalized bosonic light mode, such as provided by an optical resonator, resembles a quantized transverse magnetic field of tunable strength. As a consequence, the symmetry-broken magnetic state breaks down for strong enough light-mater interactions to a paramagnetic state.
The non-local character of the bosonic mode can change the quantum phase transition in a drastic manner, which we analyze quantitatively for the simplest case of a chain geometry (Dicke-Ising chain). The results show a direct transition between a magnetically ordered phase with zero photon density and a magnetically polarized phase with lasing behaviour of the light. Our predictions  are equally valid for the dual quantized Ising chain in a conventional transverse magnetic field.
\end{abstract}

\maketitle

\section{Introduction}
The investigation of quantum critical behaviour in correlated quantum many-body systems is an active research field over many decades in condensed matter physics; intriguing universal behavior close to quantum critical points gives rise to many fascinating quantum materials with exciting collective effects. Indeed, such physics can be understood in terms of universality classes, which are only dependent on dimension and the underlying symmetry of the system. As a consequence, the quantum critical properties of quantum matter can be described by paradigmatic models for each universality class, which in many cases correspond to interacting quantum spin systems. The most paradigmatic model in this context is the nearest-neighbor Ising model \cite{Book_Ising_2013}.

In contrast, in standard quantum optics setups, the goal is to understand and exploit the influence of (strong) light-matter interactions on a collection of non-interacting matter entities such as spins, atoms, or molecules. Such interactions are obtained by an increase in the optical density of modes around electronic resonances as it is typically done in cavity quantum electrodynamics \cite{haroche1989cavity,berman1994cavity,walther2006cavity}. The non-local character of the interaction gives rise to an effective, typically long-range coupling, between the matter degrees of freedom leading to many interesting physical effects. For example, cavity-induced long range interactions in a quantum degenerate gas inside an optical resonator can give rise to non-equilibrium quantum phase transitions as well as to novel quantum phases such as supersolids or spin glasses \cite{ritsch2013cold}. Among other models, the simplest one typically studied is the quantum Rabi (or Dicke) Hamiltonian where isolated quantum spins are coupled to a quantum light mode~\cite{ritsch2013cold}.

It is then a natural next step to investigate the interplay between strong matter-matter and strong light-matter interactions. From a condensed matter perspective one might expect to tune the properties of quantum materials by quantum light as recently theoretically discussed~\cite{mazza2019superradiant,martin2019manipulating,xiao2019cavity,kiffner2019mott,sentef2020}. From a quantum optics perspective one might aim at engineering interesting novel facets of quantum light which exploits the mapping of the intrinsic matter-matter corelations onto photons. On the experimental side, this research direction is currently actively pursued especially towards engineering novel materials with optimized functionality such as enhanced charge/energy transfer and transport in organic semiconductors~\cite{orgiu2015conductivity,zhong2016non,zhong2017energy}, modified chemical reactivity~\cite{hutchinson2012modifying,schwartz2013polariton} or modified superconducting transition temperatures~\cite{thomas2019exploring}. Recent theoretical~\cite{Mivehvar2017disorder} and experimental~\cite{Kroeze2018spinor} endeavors with quantum gases have shown the occurrence of non-equilibrium phase transitions which are characterized by spinor self-ordering in the presence of quantum field driving.

Here we approach this interesting physical domain by adding the above mentioned most paradigmatic models for matter-matter and for light-matter interaction, namely
the nearest-neighbor Ising model and a quantum Rabi Hamiltonian. We show that the quantum Rabi Hamiltonian corresponds to a {\it quantized} transverse magnetic field, so that we dub the full system quantized transverse-field Ising model (QTFIM). Indeed, for a finite density of photonic states, the quantum Rabi coupling reduces essentially to a classical transverse magnetic field while the discrete quantal character of the light becomes essential when the photon number is small and the coupling is large. It is then possible to tune a zero-temperature quantum phase transition by varying the strength of the quantized transverse magnetic field stemming from the light-matter interaction. For weak fields, the system realizes a magnetically ordered symmetry-broken phase with a small number of photons. In the strong-coupling limit the matter system is a quantum paramagnet while the light component reaches a lasing regime well described by a coherent state.

This general behaviour is qualitatively and quantitatively analyzed here for a QTFIM on a one-dimensional chain. The relevance of this model has been proposed for a variety of experimental platforms~\cite{Lee04} and a direct implementation within circuit QED has been shown~\cite{Zhang14}.
We find that the quantum phase transition is drastically altered by the quantized nature of the transverse field. While the conventional transverse-field Ising chain is exactly solvable \cite{Pfeuty} and known to be in the 2D Ising universality class, in contrast, the QTFIM chain possesses a $\mathbb{Z}_2\times\mathbb{Z}_2$ symmetry so that the phase transition becomes first-order \cite{Zhang14} between two symmetry-broken phases. Here, we determine this phase transition quantitatively by exploiting the quantized nature of the field. In the thermodynamic limit, a perturbative treatment in the weak coupling regime indicates that the ground-state energy per site of the bare Ising model is unchanged to any order. In the strong-coupling lasing limit the results are shown to exactly correspond to the ones predicted by the conventional transverse-field Ising model (TFIM) of the high-field polarized phase. This is confirmed by numerical calculations for a finite number of spins. Furthermore, we extend a well-known duality for the TFIM to the QTFIM, which results in a quantized Ising chain in a conventional transverse magnetic field.

The article is organized as follows. In Sect.~\ref{sect::model} we introduce the microscopic model we focus on in this work. In Sect.~\ref{sect::limits} we discuss the most important limiting cases of the QTFIM. Afterwards, in Sect.~\ref{sect::QTFIM}, we present our analytical calculations in the weak- and strong-coupling regime for the QTFIM on a one-dimensional chain. The analytical findings are combined with numerical diagonalizations for small systems in order to discuss the phase diagram of the QTFIM, which is done in Sect.~\ref{sec::phase_diagram}. In the final Sect.~\ref{sect::discussion} we conclude and elaborate on potential experimental realizations.

\section{Model}
\label{sect::model}
We investigate the QTFIM being the sum of an Ising interaction and a quantum Rabi Hamiltonian
\begin{equation}
  \label{eq::full_H}
 \mathcal{H}_{\rm QTFIM}= \mathcal{H}_{\rm Ising} + \mathcal{H}_{\rm Rabi} ,
\end{equation}
where the two interactions are expressed in terms of collective spin operators $\hat{S}_\alpha$ and a bosonic mode with annihilation (creation) operator $\hat{a}$ ($\hat{a}^{\dagger}$) and read
\begin{eqnarray}
 \mathcal{H_{\rm Ising}}&=& -J\sum_{\langle i,j\rangle} \sigma^z_{i}\sigma^z_{j}\\
 \mathcal{H_{\rm Rabi}}&=&  \omega_0 \; \hat{S}_z^{\phantom{\dagger}}+\frac{g}{\sqrt{N}} \left( \hat{a}^\dagger+\hat{a}^{\phantom{\dagger}}\hspace*{-1mm}\right) \hat{S}_x + \omega_{\rm c}\;\hat{a}^\dagger\hat{a}^{\phantom{\dagger}}\, .
\label{eq::specific_H}
\end{eqnarray}
The collective spin operators $\hat{S}_\alpha$ with $\alpha\in\{x, y, z\}$ are defined as $\hat{S}_\alpha\equiv \textstyle \sum_j\sigma^{\alpha}_{j}/2$
in terms of Pauli matrices satisfying \mbox{$[\sigma^{\alpha},\sigma^{\beta}]=2\mathrm{i}\epsilon_{\alpha \beta \gamma}\sigma^\gamma$} such that \mbox{$[\hat{S}_{\alpha},\hat{S}_{\beta}]=\mathrm{i}\epsilon_{\alpha \beta \gamma}\hat{S}_\gamma$}.
We denote the eigenbasis of $\hat{S}_x$ by $|{m_j}\rangle$ where the index \mbox{$j=1,\dots,2^N$} spans the whole Hilbert space and the possible values of $m_j$ are in the range \mbox{$\{-N/2, -N/2+1, \ldots, N/2\}$}.
Notice that each state $|m_j\rangle$ is ${N \choose m_j + N/2}$-fold degenerate.
Creation and annihilation operators are introduced as $\hat{S}_\pm = \hat{S}_x + \mathrm{i}\, \hat{S}_y$.
A system with $J>0$ ($J<0$) is called (anti-)ferromagnetic.
The expectation value $M_z := \langle \hat{S}_z \rangle/N \in [-1/2, +1/2]$ is referred to as magnetization and represents the magnetic order parameter in the ferromagnetic case. For antiferromagnetic Ising interaction the staggered magnetization $M_z^{\rm s} := \langle \sum_i (-1)^i \sigma_z^{(i)} \rangle/2N$ is the appropriate order parameter.

The bosonic operators satisfy \mbox{$[\hat{a},\hat{a}^\dagger]=1$} and describe a confined light mode at frequency $\omega_c$. The photon-spin coupling $g$ is obtained from the collective interaction of all spins with the bosonic mode and depends on the optical confinement (or equivalently, on the density of available optical states around the spin transition frequency). A relevant quantity and the proper order parameter for the light part of the QTFIM is the normalized photon number given by $n^{\rm ph} := \langle \hat{a}^\dagger\hat{a}^{\phantom{\dagger}} \rangle/N$.

The QTFIM possesses a $\mathbb{Z}_2\times\mathbb{Z}_2$ symmetry ($\mathbb{Z}_2$ symmetry) for $\omega_0=0$ ($\omega_0\neq0$). The first symmetry refers to the spin-flip symmetry of the Ising model, which is present only for $\omega_0=0$. In the pure transverse-field Ising chain this symmetry gives rise to a second-order phase transition in the $2D$ Ising universality class between the symmetry-unbroken polarized phase and a symmetry-broken ordered phase with finite magnetization $M_z$ as order parameter. The second symmetry is already present in the pure  quantum Rabi (Dicke) Hamiltonian ($J=0$) and triggers the second-order superradiant quantum phase transition separating the normal phase and the symmetry-broken lasing phase with finite photon density $n^{\rm ph}$. Indeed, the combined transformation $\hat{S}_x\rightarrow -\hat{S}_x$ and $\hat{a}\rightarrow -\hat{a}$ leaves the full QTFIM invariant.

With the notation $\alpha=g/(\omega_c\sqrt{N})$ one can define a generalized (conditional) displacement operator \mbox{$\hat{D} :=\exp\left\{\alpha \hat{S}_x(\hat{a}^\dagger - \hat{a})\right\}$}. This operator displaces a vacuum state into a coherent state in the photon subspace with an amplitude conditioned on the value of the $S_x$-operator in the Hilbert space of the spins. The transformation diagonalizes the quantum Rabi Hamiltonian for the case $\omega_0=0$ and transforms the Ising interaction as follows (see Appendix for derivation)
\begin{widetext}
	\begin{subequations}
\begin{align}
	\hat{D}\mathcal{H}_{\rm Rabi}\hat{D}^\dagger &=\omega_c \left[\hat{a}^\dagger \hat{a}^{\phantom\dagger} - \alpha^2\hat{S}_x^2\right]+ \omega_0 \left[\hat{S}_z \cosh\left[ \alpha \left(\hat{a}^\dagger - \hat{a}\right)\right] + \mathrm{i}\, \hat{S}_y \sinh\left[\alpha \left(\hat{a}^\dagger - \hat{a}\right)\right]\right]\, \label{diagRabi} \\
    \hat{D} \mathcal{H}_{\rm Ising}\hat{D}^\dagger &=\sum_{\langle i,j\rangle} \sigma^z_i\sigma^z_j-\frac{\mathrm{i}}{2}\left[\sum_{\langle i,j\rangle} \left(\sigma^y_i\sigma^z_j+\sigma^z_i\sigma^y_j\right)\right] \sinh\left[2\alpha \left(\hat{a}^\dagger - \hat{a}\right)\right]+\frac{1}{2}\left[\sum_{\langle i,j\rangle} \left(\sigma^z_i\sigma^z_j+\sigma^y_i\sigma^y_j\right)\right] \left(\cosh\left[2\alpha \left(\hat{a}^\dagger - \hat{a}\right)\right]-1\right)\,
	\label{TransformedH}
\end{align}
	\end{subequations}
  \end{widetext}
Interestingly, the displaced Ising interaction still contains a conventional nearest-neighbor Ising interaction extensively scaling with the number of spins, which is not coupled to the photonic operators. The other two terms are more complex representing different types of nearest-neighbor spin interactions coupled to highly non-linear photonic operators. The latter do not contain contributions of the order $\alpha^0$, which we will use in Subsect.~\ref{sssect::las_phase} to simplify the perturbation theory in the strong-coupling limit.

\begin{figure*}
	\centering
	\begin{minipage}{0.80\columnwidth}
		\centering
		\includegraphics[width=\textwidth]{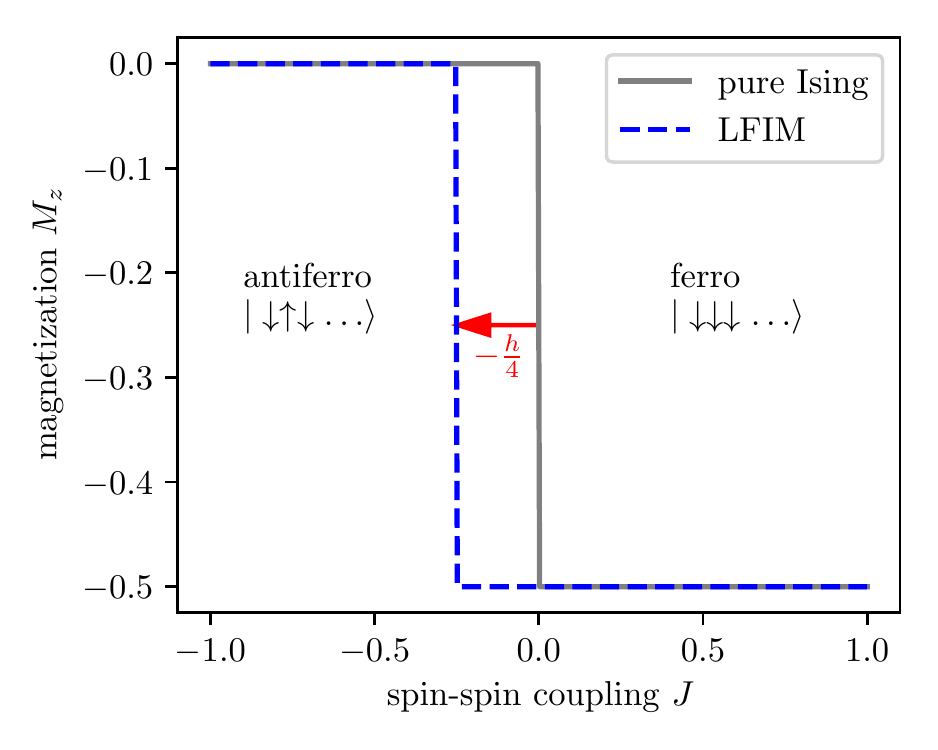}\\
	\end{minipage}
	\begin{minipage}{0.80\columnwidth}
		\centering
		\includegraphics[width=\textwidth]{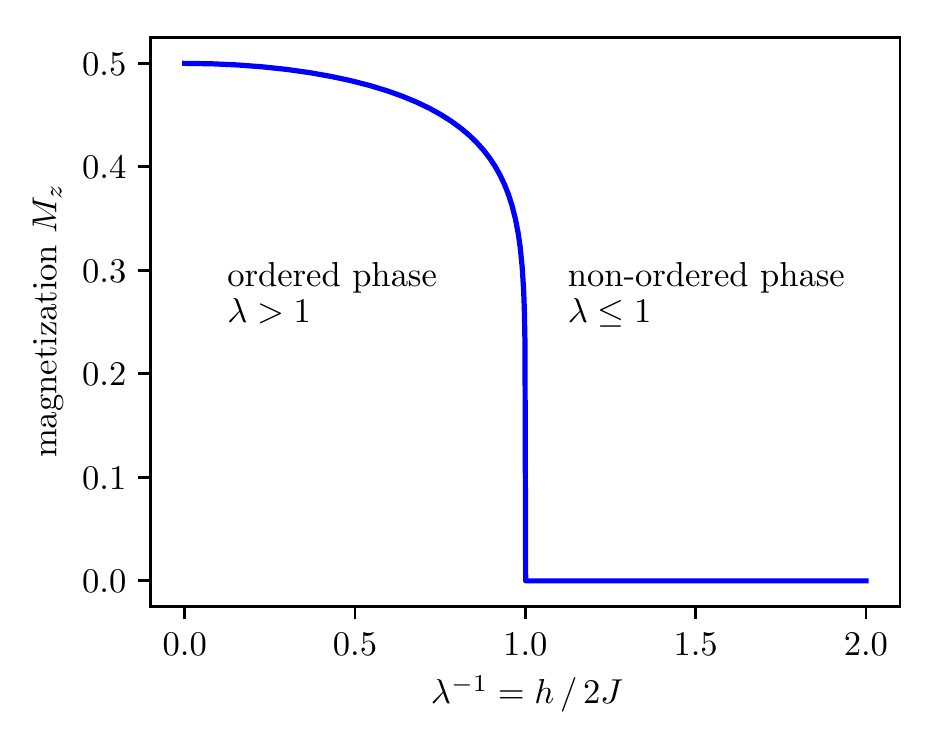}\\
	\end{minipage}
	\caption{{\it Left}: Ground-state magnetization $M_z$ of the pure Ising model and the LFIM as a function of the spin-spin coupling $J$. The change from the antiferromagnetic to the ferromagnetic ground state is at $J = 0$ ($J = -h/4$) for the pure Ising model (for the LFIM). {\it Right}: Magnetization $M_z$ for the TFIM as a function of $\lambda^{-1} = h/2J$. For small values of $h/2J$ the system is magnetically ordered, i.e., the eigenstate approaches the ferromagnetic state $|\Uparrow\rangle$. For large values of $h/2J$ the system approaches the polarized state $|\Rightarrow\rangle\equiv|\rightarrow\rightarrow\rightarrow\ldots\rangle$. By replacing the magnetization $M_z$ by the staggered magnetization $M_z^{\rm s}$ and $J$ by $-J$, the plot is also valid for an antiferromagnetic Ising interaction with $-J>0$. In this case the order approaches the state $|\Downarrow\rangle$ for $-J \gg h$.}
\label{fig1}
\end{figure*}

\section{Limiting cases of the QTFIM}
\label{sect::limits}
In the following we discuss several limiting cases of the QTFIM: the bare Ising model, the pure quantum Rabi Hamiltonian, and the transverse-field Ising model.

\subsection{Ising model}
\label{subsect::Ising}
In the absence of the quantum Rabi Hamiltonian, i.e., \mbox{$\omega_0=\omega_{\rm c}=g=0$}, the QTFIM reduces to the conventional nearest-neighbor Ising model. For ferromagnetic Ising interaction with $J>0$ the model is unfrustrated on all lattice topologies so that there are two ferromagnetic ground states $|\Uparrow\rangle\equiv|\uparrow\ldots\uparrow\rangle$ and $|\Downarrow\rangle\equiv |\downarrow\ldots\downarrow\rangle$. These two states are related by the $\mathbb{Z}_2$ spin-flip symmetry which is an exact global symmetry of the Ising model. In contrast, for an antiferromagnetic Ising interaction with $J<0$ the physical properties depend strongly on the underlying lattice. Bipartite lattices like the one-dimensional chain or the two-dimensional square lattice can be mapped exactly to the ferromagnetic case by an sublattice spin rotation about the $x$-axis: $\sigma_x\rightarrow \sigma_x$, \mbox{$\sigma_y\rightarrow -\sigma_y$}, and $\sigma_z\rightarrow -\sigma_z$ on one of the two sublattices. These cases are therefore also unfrustrated. However, lattices with loops of odd length like the triangular or the kagome lattice are highly frustrated and possess an extensive number of ground states and remain classicaly disordered even at zero temperature \cite{Book_Frustrated_Ising_1986}.

The simplest case is the one-dimensional Ising chain, which we will focus on in most parts of this work. Here, for the antiferromagnetic case $J<0$, the ground-state magnetization is $M_z = 0$ while, for a ferromagnetic chain with $J>0$, the ground-state magnetization of the fully polarized state $|\Downarrow\rangle$ is $M_z = -1/2$. By introducing additionally a small longitudinal magnetic field $h \, \hat{S}_z$ with $h>0$ one obtains the longitudinal field Ising model (LFIM)
\begin{equation}
	\mathcal{H}_{\rm LFIM} = h \, \hat{S}_z - J \sum_{i,j} \sigma^z_{i}\sigma^z_{j} 
\end{equation}
and the degeneracy of the ground state is lifted. The ground state is $|\Downarrow\rangle$ for $J>-h/4$ and $|\downarrow\uparrow\downarrow\ldots\rangle$ for $J<-h/4$. Hence, as for the pure Ising model, the magnetization is again a step function where the transition is shifted towards $J=-h/4$ as shown in Fig.~\ref{fig1}.

Due to the longitudinal field the $\mathbb{Z}_2$ spin-flip symmetry is absent for the LFIM in contrast to the pure Ising model. Hence, there is no spontaneous symmetry breaking as for the TFIM discussed in the next subsection and the change of magnetization signals a first-order phase transition.
\subsection{Transverse-field Ising model}
In the limit of $\omega_0=0$ and large number of photons, the QTFIM reduces to the conventional TFIM
\begin{equation}
 \mathcal{H}_{\rm TFIM} = h \hat{S}_x - J\sum_{i,j} \sigma^z_{i}\sigma^z_{j} ,
\label{eq::transIsing_H}
\end{equation}
where $h=g^2/\omega_c$. The mapping is achieved under the assumption that the photonic subsystem is in a coherent state, allowing the replacement of photonic creation and annihilation operators by their expection values (see Appendix). As for the pure Ising model, the TFIM possesses the exact $\mathbb{Z}_2$ spin-flip symmetry $\sigma_z^{(i)}\rightarrow -\sigma_z^{(i)}$ on all sites with index $i$.

In the case of a ferromagnetic Ising interaction $J>0$ the TFIM is unfrustrated and realizes a zero-temperature phase transition between a quantum paramagnet and a $\mathbb{Z}_2$-symmetry-broken phase for any lattice in any spatial dimension $d$ \cite{Coester_2016}. The corresponding universality class is the one of the classical Ising model in dimension $d+1$. The same behaviour is also found for an antiferromagnetic Ising interaction $J<0$ on bipartite lattices, which can be mapped exactly to the ferromagnetic case by a sublattice rotation as already explained for the pure Ising model in Subsect.~\ref{subsect::Ising}.

The situation becomes more complicated in the presence of geometric frustration where different types of quantum-critical behavior as well as exotic states of quantum matter are known to occur \cite{Moessner2001}. Important examples in the framework of fully-frustrated TFIMs are the antiferromagnetic TFIM on the triangular and pyrochlore lattice. For the triangular TFIM an order by disorder mechanism \cite{Kano_1953,Villain_1980,Shender_1982} gives rise to a ground state where translational symmetry is broken and the universality class of the quantum phase transition is $3D$-XY \cite{Blankschtein1984,Moessner2001,Moessner2003,Powalski2013}. In contrast, on the pyrochlore lattice, disorder by disorder leads to a quantum-disordered Coulomb phase \cite{Hermele_2004,Shannon_2012} in the antiferromagnetic TFIM displaying emergent quantum electrodynamics and the quantum phase transition to the high-field quantum paramagnet is first order \cite{Roechner2016}.

The only exactly solvable case is the ferromagnetic (and antiferromagnetic) TFIM on a one-dimensional chain. Here, a Jordan-Wigner transformation allows to map the TFIM to a quadratic fermionic Hamiltonian which can be diagonalized by Fourier and Bogoliubov transformations~\cite{Pfeuty}.
With the exact magnetization for $h, J > 0$ and $\lambda := 2J/h$ (see also Fig.~\ref{fig1})
\begin{equation}
	M_z = \left\{
	\begin{array}{ll}
		\frac{1}{2} \left( 1 - \lambda^{-2} \right)^\frac{1}{8} & \lambda > 1\\
		0 & \lambda \le 1
	\end{array}\right.
\end{equation}
one can easily obtain the exact quantum critical point which is given by \mbox{$\lambda_{\rm c}\equiv (2J/h)_{\rm c}=1$}.
Furthermore, it is straighforward to obtain the critical exponents which correspond to the ones of the classical $2D$ Ising model, e.g., the magnetization $M_z$ (or staggered magnetization $M_z^{\rm s}$ for an antiferromagnetic Ising interaction) scales as $(\lambda-\lambda_{\rm c})^\beta$ with $\beta=1/8$ close to the critical point representing the order parameter of the system.
For $J \ll h$ the ground state is nearly the polarized state $|\Rightarrow\rangle$ whereas the ordered states $|\Uparrow\rangle$ and $|\Downarrow\rangle$ are approached for $J \gg h$. Finally, we give the explicit analytic expression of the ground-state energy per site
\begin{equation}
\label{eq::TFIM_e0_exact}
e_{0,\rm TFIM}(\lambda) = -\frac{J}{\lambda} \frac{1}{2\pi}\int_{0}^{2\pi}dk\sqrt{1+\lambda^2+2\lambda\cos(k)},
\end{equation}
which we will use in Subsect.~\ref{sssect::las_phase} for the strong-coupling perturbation theory.

\subsection{Quantum Rabi Hamiltonian}
\label{sec::quantum_rabi}
The limit of vanishing Ising interaction $J=0$ corresponds to the quantum Rabi Hamiltonian where non-interacting spins 1/2 are collectively coupled via their total spin $x$-component to a single bosonic light mode described by creation and annihilation operators $\hat{a}^\dagger$ and $\hat{a}^{\phantom{\dagger}}\hspace*{-1mm}$, respectively.

While the model is generally unsolvable, some limiting cases are analytically tractable among which i) the case of degeneracy with $\omega_0=0$, ii) the weak-coupling case with $g\ll\omega_c$ where the model is known as the Jaynes-Tavis-Cummings model \cite{jaynes1963comparison}, and iii) the Dicke Hamiltonian in the limit of an infinite number of spins $N \rightarrow \infty$ \cite{hepp1973superradiant,brankov_1975,BOGOLUBOV1976163}.
For the specific case $\omega_0=0$ the energies are at least twofold degenerate so that we dub this limit the degenerate Rabi Hamiltonian.
This is a consequence of $[\hat{S}_x,\mathcal{H_{\rm Rabi}}]=0$ so that the eigenvalues of $\hat{S}_x$ are conserved quantities.
The system therefore factorizes in a spin and in a photonic part. With the conditional displacement operator previously defined $\hat{D} := \exp{\alpha \hat{S}_x\, (\hat{a}^\dagger - \hat{a})}$ and the property $\hat{D}^\dagger \hat{a} \hat{D} = \hat{a} + \alpha \hat{S}_x$, the degenerate Rabi Hamiltonian can be transformed to the diagonal expression in Eq.~\eqref{diagRabi} (after setting $\omega_0=0$).
It is then straightforward to obtain the energies
\begin{equation}
	E_{n,j} = \omega_c \left[n - \alpha^2 m_j^2\right]
\end{equation}
with $m_j \in \{-N/2, -N/2+1, \ldots, N/2\}$.
The corresponding eigenstates of $\mathcal{H_{\rm Rabi}}$ are given by the application of $\hat{D}^\dagger$ on the eigenstates $|n\rangle\otimes|m_j\rangle$ of the transformed Hamiltonian.
Hence, the exact ground state has the quantum numbers $n=0$ and $m_j = \pm N/2$ and is obtained by the action of $\hat{D}^\dagger$ onto the ket $\left(|0\rangle \otimes |\pm N/2\rangle\right) $ leading to
\begin{equation}
\begin{aligned}
|\psi_0\rangle = \left( e^{-\frac{|\alpha N|^2}{8}}\sum_{n=0}^\infty \frac{(\mp \alpha N)^n}{2^n\sqrt{n!}}|n \rangle\right) \otimes |\pm N/2\rangle \quad ,
\label{eq::rabi_limit_w0}
\end{aligned}
\end{equation}
where $|\pm N/2 \rangle$ are eigenstates of $\hat{S}_x$ and $|n \rangle$ denotes a photonic state with $n$ photons. The ground state has therefore a Poissonian photon distribution.

\begin{figure}[t]
	\centering
	\includegraphics[width=0.8\columnwidth]{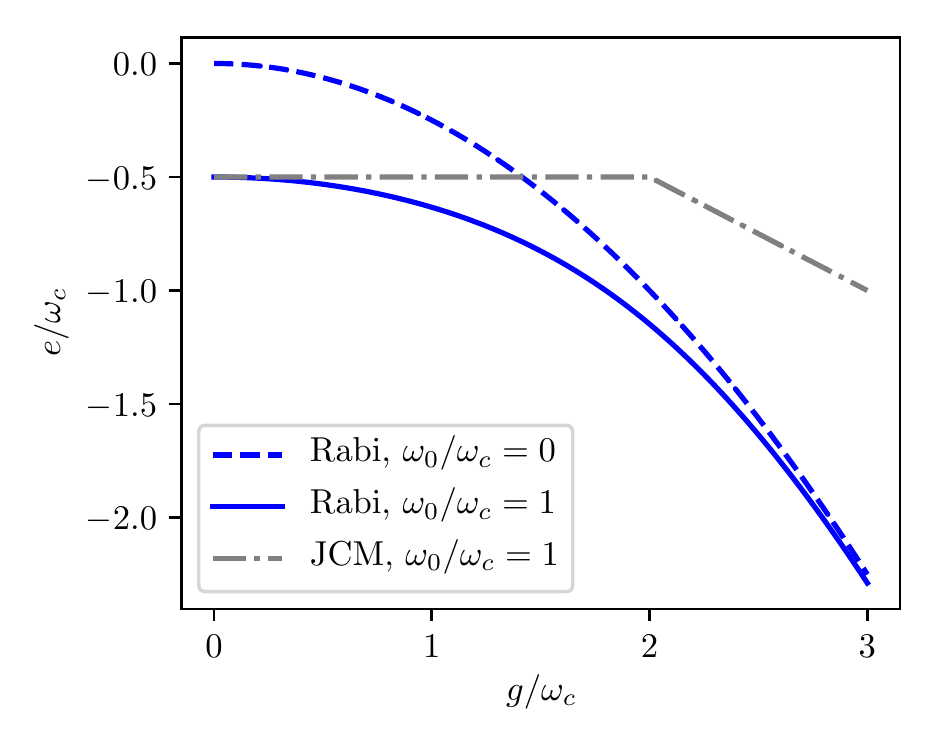}
	\caption{Ground-state energy per site $e$ for i) Rabi model for $\omega_0 \in \{0, 1\}$ and ii) Jaynes-Cummings model (JCM). In both cases a single spin is considered. For $g \ll \omega_c$, the JCM is a good approximation of the Rabi model with finite $\omega_0$ whereas for large $g \gg \omega_c$ the solution to the Rabi model with $\omega_0 = 0$ approaches the ground-state energy of the Rabi model for $\omega_0/\omega_c = 1$.}
	\label{fig2}
\end{figure}

In the general case $\omega_0\neq 0$ the Hamiltonian has to be diagonalized numerically.
Due to $[\hat{S}_x, \hat{S}_z] \neq 0$, the ground state is a complex, entangled state.
However, for large values of $g$, the degenerate Rabi model ($\omega_0 = 0$) is a good approximation as it can be seen in Fig.~\ref{fig2}.
The ground-state energy for $\omega_0 = \omega_c$ converges towards the ground-state energy for $\omega_0 = 0$ as $g$ is increased and the photon distribution approaches a Poissonian distribution.

In the case where the spin-photon system is quasi-resonant $\omega \approx \omega_c$ and the coupling is very weak $g\ll \omega_0,\omega_{\rm c}$, one can perform a rotating wave approximation where energy non-conserving terms such as $\hat{a}\hat{S}_-$ are dropped out.
Since $\hat{S}_x = (\hat{S}_+ + \hat{S}_-)/2$, the remaining Hamiltonian is
\begin{equation}
	\mathcal{H_{\rm JCM}}= \omega_0\; \hat{S}_z^{\phantom{\dagger}}+\omega_c \alpha\left[ \hat{a}\hat{S}_- + \hat{a}^\dagger\hat{S}_+\right] + \omega_{\rm c}\;\hat{a}^\dagger\hat{a}^{\phantom{\dagger}}\, .
\end{equation}
This model is also known as the Jaynes-Tavis-Cummings model.
The light-matter interaction describes the transition from the higher energy level to the lower one by annihilating a photon and the other way around.
The total number of excitations, i.e.~ the number of spins in state $|\uparrow\rangle$ plus the number of photons is always conserved, hence $[\mathcal{H}_{\rm JCM}, \omega_c\,(\hat{S}_z + \hat{a}^\dagger\hat{a})] = 0$ for any value of $\omega_0$.
Consequently, an eigenbasis exists such that the representation of the Hamiltonian reduces to a block-diagonal matrix. Then, for $N$ spins, the problem reduces to the diagonalization of $2^N \times 2^N$-matrices.
By exploiting symmetries, one can reduce the dimension of the block matrices further and obtain an analytical solution. As illustrated in Fig.~\ref{fig2},
the JCM approximates well the Rabi model for $g \ll \omega_0 = \omega_c$. For more general sets of parameters, a full check is however necessary to assess whether the rotating wave approximation is a valid simplification.

\section{QTFIM: Analytical considerations}
\label{sect::QTFIM}

The full QTFIM given in Eq.~\eqref{eq::full_H} is now investigated for the case of a one-dimensional chain of $N$ spins and nearest-neighbor Ising interactions. For the specific case $\omega_0 = 0$, which we focus on in the following, this model is a modification of the TFIM, since the magnetic field $h$ is replaced by the \emph{quantized} transverse field $g/\sqrt{N} ( \hat{a}^\dagger + \hat{a})$. We first discuss the weak-coupling limit in the magnetically ordered phase where the light-matter interaction is assumed to be small. Afterwards, we turn to a perturbative treatment of the strong-coupling lasing phase.
\subsection{Weak-coupling limit: magnetically-ordered phase}
\label{sssect::mag_ordered}
We consider the weak-coupling limit $g\ll J,\omega_{\rm c}$ for $\omega_0=0$, where the system is magnetically ordered, i.e.~the $\mathbb{Z}_2$ spin-flip symmetry is spontaneously broken, and the density of photons is zero. The unperturbed Hamiltonian is therefore given as
\begin{equation}
 \mathcal{H}_0=-J\sum_{i} \sigma^z_{i}\sigma^z_{i+1}+\omega_{\rm c}\;\hat{a}^\dagger\hat{a}^{\phantom{\dagger}}
\end{equation}
and the perturbation reads
\begin{equation}
 \mathcal{V}=\frac{g}{\sqrt{N}} \left( \hat{a}^\dagger+\hat{a}^{\phantom{\dagger}}\hspace*{-1mm}\right) \hat{S}_x\, .
\end{equation}
Here we calculate the corrections to the ground-state energy in powers of $g$. To this end we choose the symmetry broken magnetic state $|\Uparrow\rangle\equiv |\uparrow\ldots\uparrow\rangle$ as one of the two Ising ground states so that the unperturbed ground state of the full model corresponds to
\begin{eqnarray}
 |\psi_0\rangle &=& \ket{0}\otimes\ket{\Uparrow}\, .
\end{eqnarray}
The associated unperturbed ground-state energy is given by $E_{0,{\rm weak}}^{(0)}=-JN$. Clearly, the first-order correction to the ground-state energy vanishes, since the light-matter coupling always changes the photon number by one. In second order, one obtains
\begin{eqnarray}
 E_{0,{\rm weak}}^{(2)} &=& \bra{\psi_0}\mathcal{V}\,\frac{1}{E_0^{(0)}-\mathcal{H}_0}\,\mathcal{V}\ket{\psi_0}= -\frac{1}{4}\frac{g^2}{2J+\omega_c}\, .
\end{eqnarray}
The energy correction is therefore {\it not} extensive.
The same is true in any order of perturbation theory.
Indeed, since the perturbation scales as $g/\sqrt{N}$, one gets a suppressing factor $N^{-k/2}$ in order $k$ of perturbation theory which cannot be compensated by the appearing sums in the general perturbative expressions in order to yield an extensive contribution.
As a consequence, in the thermodynamic limit $N\rightarrow\infty$, the ground-state energy per site is exactly
\begin{equation}
    e_{0, \mathrm{weak}} = \frac{E_{0, \mathrm{weak}}}{N} = \frac{E^{(0)}_{0, \mathrm{weak}}}{N} = -J .
\end{equation}
Calculating the first-order correction to the ground-state vector, one obtains
\begin{equation}
  |\psi_0^{(1)}\rangle = -\frac{g}{2 \sqrt{N}}  \frac{1}{2J + \omega_c} \ket{1} \otimes \sum_\xi |\xi\rangle
\end{equation}
where
\begin{equation}
  |\xi\rangle := |\uparrow\ldots\uparrow\underset{\xi}{\downarrow}\uparrow\ldots\rangle.
\end{equation}
Consequently, the magnetization is not altered in leading order in the thermodynamic limit, which is true to any order in perturbation theory.

\subsection{Strong-coupling limit: lasing phase}
\label{sssect::las_phase}
Next we investigate the strong-coupling limit for $\omega_0=0$ where the Ising interaction is treated as a perturbation to the Rabi Hamiltonian. We calculate the expression for the ground-state energy per site as well as the corrected ground-state wave vector from which the corresponding photon number statistics can be deduced. For both, energy and the ground state, we start by presenting the derivation which employs the displaced basis. However, in both cases the same results are also obtained in the bare basis, which we show explicitly for the ground-state wave function. Furthermore, for the ground-state energy, again two distinct approaches are pursued: i) a mapping onto the TFIM after ignoring negligible non-extensive contributions in the displaced Hamiltonian and ii) a mean-field approach based on the linearization around large classical amplitudes of the photon field under the condition $g^2/\omega_c\gg1$ in the initial basis. Both approaches yield the same results and provide an elegant mapping of the QTFIM onto the TFIM which admits an exact solution for the chain geometry.

We first analyze the perturbation term in Eq.~\eqref{TransformedH} in the displaced basis as modification of the eigenenergies of the displaced diagonalized Rabi part of Eq.~\eqref{diagRabi}.
The unperturbed Hamiltonian is therefore given by
\begin{equation}
	\mathcal{H}_0 =\omega_c \left[\hat{a}^\dagger \hat{a}^{\phantom\dagger} - \alpha^2\hat{S}_x^2\right]
    \label{h0_diagRabi}
  \end{equation}
with eigenstates written as
\begin{equation}
  |\psi_{n,m, l}^{(0)}\rangle = |n\rangle \otimes |m, l\rangle
\end{equation}
where $|n\rangle$ are the eigenstates of $\hat{a}^\dagger \hat{a}^{\phantom{\dagger}}$ and \mbox{$\hat{S}_x |m, l\rangle = m \, |m, l\rangle$} with an index $l$ to take into account the degeneracy of the spin states. The corresponding bare eigenenergies are
\begin{equation}
  E_{n,m,l}^{(0)} = n\omega_c - \frac{g^2}{N \omega_c}m^2 = n\omega_c - \omega_c \alpha^2 m^2
\end{equation}
independent of $l$. Apparently, the ground state is twofold degenerate with eigenstates
\begin{equation}
 \begin{aligned}
\ket{\Psi}_{\rm left} &\equiv |\psi_{0,-N/2, 1}^{(0)}\rangle = \Ket{0} \otimes \ket{\Leftarrow}\\\ket{\Psi}_{\rm right} &\equiv |\psi_{0,+N/2, 1}^{(0)}\rangle = \Ket{0} \otimes \ket{\Rightarrow}
\end{aligned}
\end{equation}
and energy $E_{0,{\rm strong}}^{(0)}\equiv E_{0,\pm N/2,1}^{(0)}=-g^2 N/4\omega_c$.
Furthermore, we define the ground-state energy per site $e_{0,{\rm strong}}^{(0)} = -g^2/4\omega_c$.
Without loss of generality we choose the eigenstate at order zero of perturbation theory as $\ket{\Psi}_{\rm right}$.

In the following subsections we show that the perturbative strong-coupling expansion of the QTFIM chain is equivalent to the high-field expansion of the conventional TFIM, first by application of the perturbation theory introduced above and second within a mean-field approach. Afterwards, we present the first-order correction to the ground state and calculate the associated ground-state photon distribution.

\subsubsection{Energetics}
\label{ssssect::energy}
We calculate the ground-state energy per site \mbox{$e_{0,{\rm strong}}=E_{0,{\rm strong}}/N$}  perturbatively in $J\omega_c/g^2$ for \mbox{$N\rightarrow \infty$} in the displaced basis.

The most important point to realize is that the last two terms on the right hand side of Eq.~\eqref{TransformedH} do not give rise to extensive corrections to the ground-state energy $E_{0,{\rm strong}}^{(0)}$ for finite perturbation orders. The reason for that is the factor $\alpha\propto 1/\sqrt{N}$ in the photonic expressions. Hence to obtain extensive perturbative corrections of the ground-state energy $E_{0,{\rm strong}}$ only the Ising interaction $\sum_i \sigma^z_i\sigma^z_{i+1}$ is relevant and the problem reduces to finding the ground-state energy of
\begin{equation}
  \label{eq::spin_ham_strong}
  - \omega_c \alpha^2\hat{S}_x^2 - J \sum_i \sigma^z_i\sigma^z_{i+1}
\end{equation}
where the light part is not written down anymore since it completely decouples from the spin part in this situation.
If one assumes that the ground state of this Hamiltonian is a superposition of states which are given by a finite number of spin flips with respect to the unperturbed ground state (which we prove in the Appendix), the $\hat{S}_x^2$ operator can be rewritten for $N \rightarrow \infty$ as follows. For a state resulting from flipping $a < \infty$ spins of $\ket{\Psi}_{r}$ the corresponding eigenenergy of $- \omega_c \alpha^2\hat{S}_x^2$ is
\begin{equation}
  -\frac{N g^2}{4 \omega_c} + a \frac{g^2}{\omega_c} -a^2\frac{g^2}{\omega_cN} \rightarrow -\frac{N g^2}{4 \omega_c} + a \frac{g^2}{\omega_c}
\end{equation}
and the spectrum is equidistant.
Therefore the unperturbed part of the Hamiltonian in Eq.~\eqref{eq::spin_ham_strong} can be replaced by an effective $\hat{S}_x$ operator and the total Hamiltonian reads
\begin{equation}
  \label{eq::simplified_qtfim_strong}
  \frac{N g^2}{4 \omega_c} + \frac{g^2}{\omega_c} \hat{S}_x - J \sum_i \sigma^z_i\sigma^z_{i+1} \, .
\end{equation}

Comparing with the conventional TFIM on a chain Eq.~\eqref{eq::transIsing_H}, one immediately sees that for $h=g^2/\omega_{\rm c}$ a perturbation in $J/h$ yields the same ground-state corrections in $\mathcal{H}_{\rm TFIM}$ as in the QTFIM.
We conclude that given the perturbation series obtained from the analytic expression for the TFIM $e_{0,\rm TFIM}(2J/h)$, the one of the QTFIM is
\begin{equation}
    e_{0,\rm strong}\left(\frac{J\omega_c}{g^2}\right)
    = e_{0,\rm TFIM}\left(\frac{2J\omega_c}{g^2}\right)-e_{0,{\rm strong}}^{(0)} \,.
\end{equation}
In regions where the perturbative expansion converges this can be expressed with the exact formula for the ground-state energy of the Ising model in a transverse field \cite{Pfeuty} given in Eq.~\eqref{eq::TFIM_e0_exact} with $\lambda = 2J\omega_c/g^2$.
The final result for $e_{0,\rm strong}\left( J\omega_c/g^2\right)$ is then
\begin{equation}
  \begin{aligned}
  e_{0,\rm strong} = \frac{g^2}{4 \omega_c} - \frac{g^2}{2 \omega_c}\frac{1}{2 \pi} \int_0^{2\pi} \mathrm{d}k \sqrt{1 + \lambda^2 + 2 \lambda \cos{(k)}}\,.
\end{aligned}
\end{equation}

The same result is obtained by a mean-field approach presented in detail in the Appendix where the photonic operators in the original Hamiltonian of Eq.~\eqref{eq::full_H} are replaced by $\langle \hat{a}^{(\dagger)}\rangle + \delta \hat{a}^{(\dagger)}$.
Discarding fluctuations of order $\sqrt{N}$ and assuming that the photonic part is the coherent state, which solves the model for $J=0$, leads to the same Hamiltonian as in Eq.~\eqref{eq::simplified_qtfim_strong}. Thus, mean-field theory in the light part is an equivalent approach to solve the model in the strong-coupling limit.

\subsubsection{Ground state and photon distribution}
\label{ssssect::state_photon}

Now we calculate the first-order correction $|\psi^{(1)}_{\rm right} \rangle$ to the unperturbed ground state $\ket{\Psi}_{\rm right}$ of the Rabi Hamiltonian for \mbox{$\omega_0=0$} due to the Ising interaction. This can be done by performing the calculation first in the displaced basis and subsequently re-transforming to the original basis as shown in the Appendix. We further show that the computation of the ground-state correction using the bare basis yields the same result.

The first-order correction of the state vector leads to
\begin{equation}
    |\psi^{(1)}_{\rm right} \rangle = \left( \sum_n f_n \hat{D}_\mathrm{ph}^\dagger(\alpha (N/2-2))|n\rangle \right) \sum_\nu |\nu \rangle
\end{equation}
where the states with two nearest-neighbor spin flips and therefore $m=(N-4)/2$ are labeled as
\begin{equation}
  \label{eq::def_nu}
  | \nu \rangle := |\rightarrow\,\rightarrow\,\ldots\,\rightarrow\,\underset{\nu}{\leftarrow}\,\underset{\nu+1}{\leftarrow}\,\rightarrow\,\ldots\rangle
\end{equation}
and the coefficients $f_n$ are given as
\begin{equation}
  f_n = \frac{e^{-2|\alpha|^2} (-2\alpha)^n}{\sqrt{n!}}\frac{1}{\omega_c \alpha^2 \left(4 - 2N \right) - n \omega_c}\, .
\end{equation}
The photonic displacement operator $\hat{D}_\mathrm{ph}(z)$ is defined as
\begin{equation}
  \hat{D}_\mathrm{ph}(z) := \exp\left(z\hat{a}^\dagger - z^*\hat{a}\right)
\end{equation}
and acts only on the photonic part of the Hilbert space. In order to calculate the photon distribution in the strong-coupling limit, we trace over the spin part of the total ground state $|\psi_{\rm right}\rangle = | \psi^{(0)}_{\rm right} \rangle + | \psi^{(1)}_{\rm right} \rangle$ in first-order perturbation theory. One then obtains the density matrix
\begin{equation}
  \begin{aligned}
  \rho &= \sum_{m,l} \langle m, l | \psi_{\rm right} \rangle \langle \psi_{\rm right} |  m, l \rangle\\
       &= | \psi^{(0)}_{{\rm right}, \mathrm{ph}} \rangle \langle \psi^{(0)}_{{\rm right}, \mathrm{ph}} | + N | \psi^{(1)}_{{\rm right}, \mathrm{ph}} \rangle \langle \psi^{(1)}_{{\rm right}, \mathrm{ph}} |
  \end{aligned}
\end{equation}
where the photonic states are defined as follows
\begin{equation}
  \begin{aligned}
    | \psi^{(0)}_{{\rm right}, \mathrm{ph}} \rangle &:= e^{-\frac{|\alpha|^2 N^2}{8}}\sum_{n = 0}^\infty \frac{(-\alpha N)^n}{2^n \sqrt{n!}} | n \rangle \\
    | \psi^{(1)}_{{\rm right}, \mathrm{ph}} \rangle &:= -J\sum_{n = 0}^\infty f_n \hat{D}_\mathrm{ph}^\dagger(\alpha\frac{N-4}{2})|n\rangle .
  \end{aligned}
\end{equation}
The photon distribution $P(n)=\Tr\left( | n \rangle \langle n |\rho \right)$ is then given by the following expression
\begin{widetext}
\begin{equation}
    P(n) = e^{-\frac{|\alpha|^2 N^2}{4}} \frac{(-\alpha N)^{2n}}{2^{2n} {n!}} +J^2 \sum_{k, k^\prime = 0}^\infty f_k\;f_{k^\prime} \cdot \langle n | \hat{D}_\mathrm{ph}^\dagger(\alpha \frac{N-4}{2}) \ket{k} \langle k^\prime | \hat{D}_\mathrm{ph}(\alpha\frac{N-4}{2}) \ket{n}\, .
\end{equation}
\end{widetext}
We stress that the first-order contribution to the ground state does not change the form of the photon distribution for $N \rightarrow \infty$.
Taking into account that $\alpha \rightarrow 0$ and \mbox{$\alpha^2 N = \text{const}$}, one can see that $ f_k = \mathcal{O}\left( N^{-k/2} \right)$
for $N \rightarrow \infty$.
In this limit, also the order of the overlap $\langle n | \hat{D}_\mathrm{ph}^\dagger(\alpha \frac{N-4}{2}) \ket{k} $ can be determined.
Therefore we use $|k\rangle \propto (\hat{a}^\dagger)^k |0\rangle$ and the commutation relation $[\hat{D}_\mathrm{ph}(x), \hat{a}^\dagger] = x \hat{D}_\mathrm{ph} (x)$ in order to express the overlap in terms of $\langle n | \hat{D}_\mathrm{ph}^\dagger(\alpha \frac{N-4}{2}) | 0 \rangle$.
Then, one can show that
\begin{equation}
  \langle n | \hat{D}_\mathrm{ph}^\dagger(\alpha \frac{N-4}{2}) \ket{k} \rangle = \mathcal{O} \left(\sqrt{N}^k\right) \langle n | \hat{D}^\dagger_\mathrm{ph}(\alpha N/2)| 0 \rangle \,.
\end{equation}
As a consequence, the ground state remains a coherent state with a Poissonian photon distribution. This is different for finite $n$, where differences between odd and even $n$ are present and the photonic part can show non-classical features.
\section{Phase diagram}
\label{sec::phase_diagram}
We are now in the position to analyze the phase diagram of the QTFIM on a chain. To this end we combine our analytical findings presented in the weak- and strong coupling limit in the last section with numerically exact diagonalizations of small systems up to $N=10$.

\begin{figure}
  \includegraphics[width=\columnwidth]{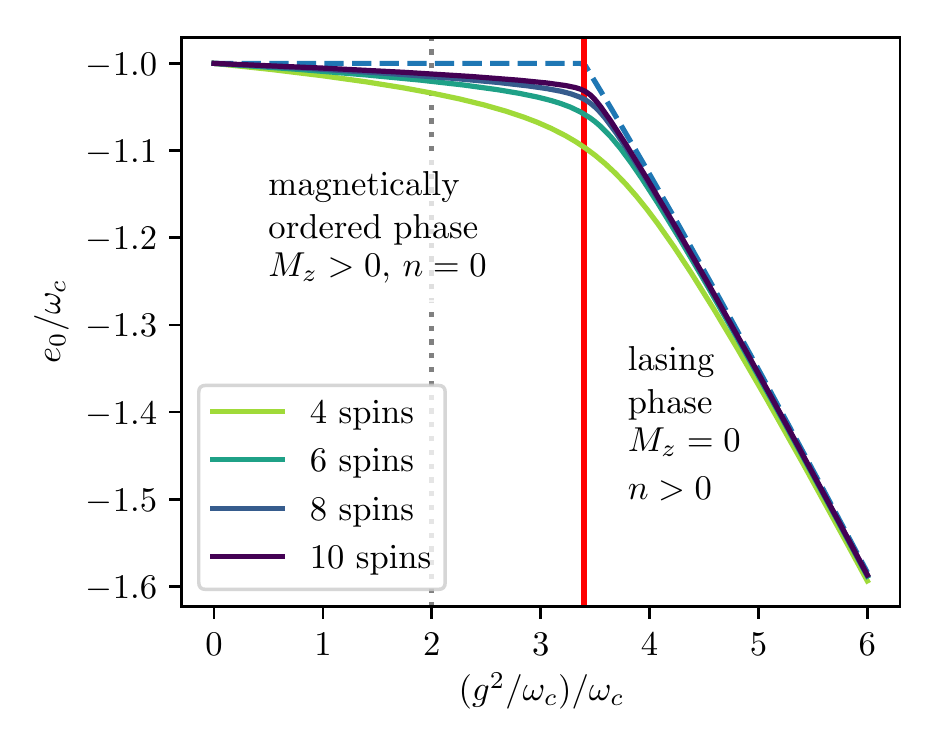}
  \caption{Ground-state energy per site $e_0$ for $J/\omega_c = 1$: Comparison of the analytical result (blue dashed line) and numerical data obtained for $N=4$ to $N=10$ spins (solid lines). The quantity $g^2/\omega_c$ serves as an analogue of the magnetic field strength of the conventional TFIM. The vertical red line indicates the location of the quantum phase transition separating the magnetically ordered phase for small $g$ from the lasing phase at large $g$.}
  \label{fig::energypersite}
\end{figure}

Most importantly, we can locate the quantum phase transition between the magnetically ordered phase and the lasing phase to an arbitrary precission in the thermodynamic limit, since i) the ground-state energy of the magnetically ordered phase does not depend on the light-matter interaction and ii) the ground-state energy of the lasing phase has been shown to correspond exactly to the one of the transverse-field Ising chain in the polarized phase. The crossing of both ground-state energies
\begin{equation}
  \frac{J_\mathrm{crit}}{\omega_c} = \beta \left(\frac{g_\mathrm{crit}}{\omega_c}\right)^2
\end{equation}
with $\beta \approx 0.294$ yields the quantum phase transition point of the QTFIM on a chain geometry. For $J/\omega_c = 1$, the analytic and numerical ground-state energies per site are shown in Fig.~\ref{fig::energypersite}. The phase transition at \mbox{$g^2/\omega_{\rm c}^2\approx 3.40$} (vertical red line) is first order, which can be seen directly from the kink of the ground-state energy in the thermodynamic limit (dashed line), which separates the magnetically ordered phase (small $g^2/\omega_{\rm c}$) and the lasing phase (large $g^2/\omega_{\rm c}$). We note that the numerical data for finite $N$ displays the same qualitative behaviour and approach the expressions of the thermodynamic limit in a monotonous fashion. Altogether, when comparing to the conventional transverse-field Ising chain, the order of the phase transition changes to first order in the QTFIM and the extension of the ordered phase becomes larger.

\begin{figure}
  \includegraphics[width=\columnwidth]{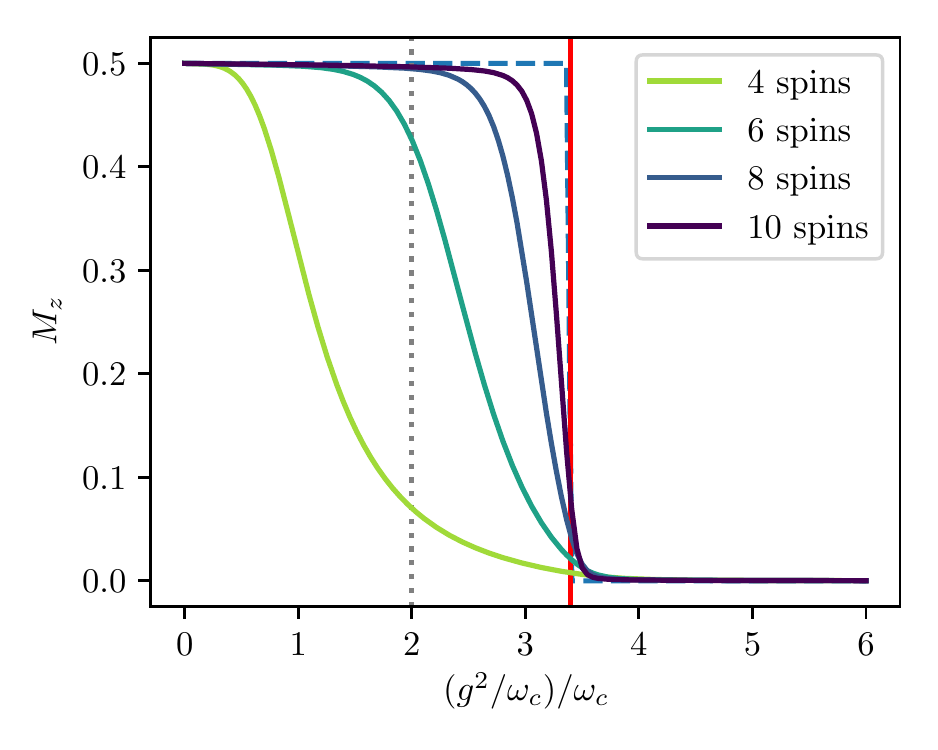}
  \caption{Magnetization per spin $M_z$ for $J/\omega_c = 1$: The analytical result (blue dashed line) is approached by the numerical data for finite system sizes with fixed $\omega_0 = 0.001\omega_c$ (solid lines). The red vertical line marks the phase transition from the magnetically ordered phase (small $g$, $M_z = 1/2$) to the lasing phase (large $g$, $M_z = 0$). The grey dotted vertical line highlights the second order phase transition of the conventional transverse-field Ising chain (see Eq.~\eqref{eq::transIsing_H}) with $h = g^2/\omega_c$.}
  \label{fig::magnpersite}
\end{figure}

\begin{figure}
  \includegraphics[width=\columnwidth]{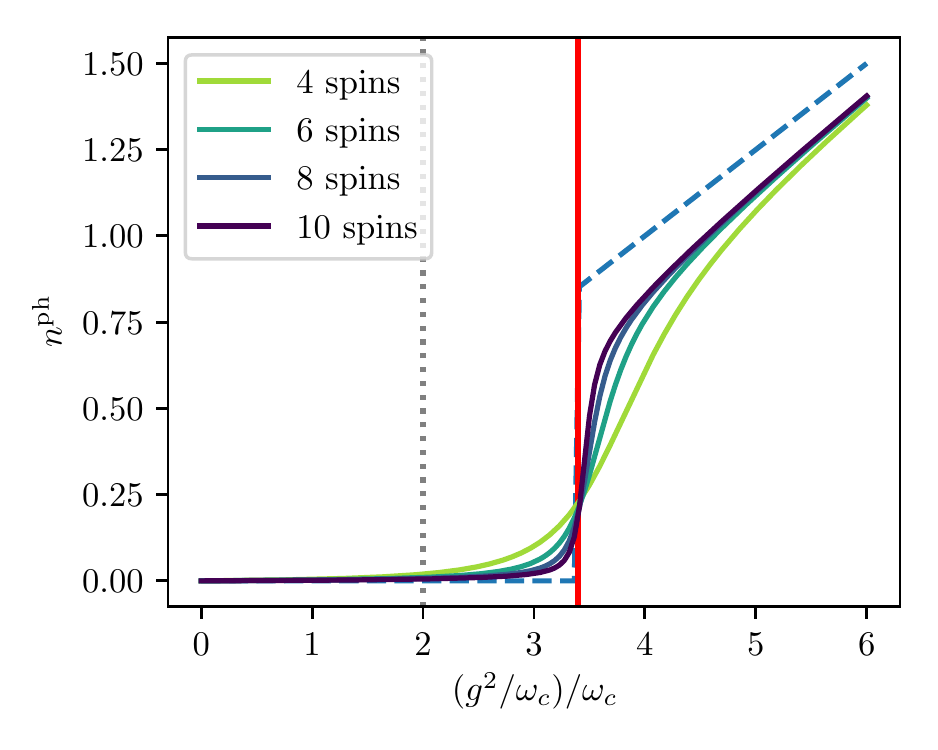}
  \caption{Photon number per spin $n$: The numerical values (solid lines) indicate the point of the transition already quite well. However, the analytical solution (dashed line) is approached very slowly with increasing $g$.}
  \label{fig::photonpersite}
\end{figure}

Both ordered phases can be characterized by an order parameter, namely the magnetization per spin $M_z$ for the magnetically ordered phase and the photon number per spin $n$ for the lasing phase. Interestingly, both order parameters can be deduced exactly in thermodynamic limit, which is again a consequence of the vanishing quantum fluctuations in the weak-coupling regime and the mapping to the conventional TFIM for the lasing phase. The corresponding analytical and numerical results for the two order parameters are shown in Figs.~\ref{fig::magnpersite} and \ref{fig::photonpersite} for $J/\omega_c = 1$.
In the thermodynamic limit, the magnetization per spin jumps at the phase transition point from the maximally ordered value $1/2$ of the pure Ising model to zero in the lasing phase, in full agreement with the first-order nature of the phase transition. In contrast, the magnetization per spin for finite $N$ is a smooth function in the regime of the magnetically ordered phase with values smaller $1/2$ signaling true quantum fluctuations in the ground state. Interestingly, $M_z$ remains almost zero in the lasing phase even for finite $N$.
The photon number per spin $n$ in the thermodynamic limit is only finite in the lasing phase and serves as the proper order parameter. At the phase transition this quantity jumps from zero to $n\approx 0.85$  and it increases linearly as a function of $g^2/\omega_{\rm c}^2$ within the lasing phase:
\begin{equation}
  n = \left\{\begin{array}{cc}
    0 & \lambda \ge \lambda_\mathrm{crit} \\
    \frac{g^2}{4\omega_c^2} & \lambda < \lambda_\mathrm{crit}
  \end{array}\right.
\end{equation}
where $\lambda = 2J\omega_c/g^2$ and $\lambda_\mathrm{crit} \approx 0.588$. Again, the results for finite $N$ are fully consistent and approach the first-order phase transition of the QTFIM smoothly for increasing $N$.

\subsection{The dual transformation: the transverse field in a quantized Ising chain}
\label{sssect::duality}
Next we introduce an exact duality of the QTFIM for $\omega_0=0$ to a transverse field in a {\it quantized} Ising chain. Such a Kramers-Wannier duality is well known for the conventional TFIM on a chain which even displays an exact self-duality. To this end we introduce pseudo-spins 1/2 on links $\nu$ described by Pauli matrices $\vec{\tau}$. The pseudo-spin state $|\uparrow\rangle$ ($|\downarrow\rangle$) is then identified with ferromagnetic (anti-ferromagnetic) spin configurations $|\uparrow\uparrow\rangle$ and $|\downarrow\downarrow\rangle$ ($|\uparrow\downarrow\rangle$ and $|\downarrow\uparrow\rangle$) on the corresponding link.

The Ising exchange is then mapped to an effective field term in the pseudo-spin language and each operator $\sigma^x_i$ in the quantum Rabi Hamiltonian becomes an effective nearest-neighbor Ising exchange $\tau^z_{\nu} \tau^z_{\nu+1}$. In total one obtains the following dual Hamiltonian of the QTFIM for $\omega_0=0$ \begin{equation}
 \mathcal{H}^{\rm QTFIM}_{\rm dual}= -J\sum_{\nu} \tau^x_{\nu} +\frac{g}{N}\left( \hat{a}^\dagger+\hat{a}^{\phantom{\dagger}}\right) \sum_{\nu}\tau^z_{\nu} \tau^z_{\nu+1} + \omega_{\rm c}\;\hat{a}^\dagger\hat{a}^{\phantom{\dagger}}\, ,
\label{eq::full_H_dual}
\end{equation}
which corresponds to a transverse field in a light-induced {\it quantized} Ising chain. Let us stress that this non-local mapping does not keep track of state properties as well as degeneracies. However, the QTFIM and its dual are isospectral and we can therefore directly conclude that also $\mathcal{H}^{\rm QTFIM}_{\rm dual}$ displays a first-order phase transition in the thermodynamic limit, but in this case between a symmetry unbroken phase with zero magnetization per spin and zero photons per spin at small $g$ and an ordered phase with finite $M_z$ {\it and} $n$ for large $g$. These findings are indeed in accordance with general considerations of Rabi Hamiltonians competing with short-range interactions \cite{DENOUDEN1976425,CAPEL1979371}.\\

\section{Conclusions}
\label{sect::discussion}
In this work we have combined two paradigmatic models, the Ising model from condensed matter physics and the quantum Rabi Hamiltonian from quantum optics. The latter corresponds to a light-induced quantized magnetic field and we therefore consider the QTFIM as the paradigmatic model to study what one could call optomagnetism.\\
Here we have investigated in detail the simplest, geometrically unfrustrated, geometry which is the one-dimensional chain and we focused on the case $\omega_0=0$. In the thermodynamic limit, the phase transition between the magnetically ordered weak-coupling phase and the strong-coupling lasing phase can be determined to an arbitrary precission. This is the consequence of the fact that quantum fluctuations are absent in the magnetic phase and that we found the appropriate connection of the lasing phase to the analytic solution of the conventional transverse-field Ising chain, both resulting from disentangling extensive and sub-extensive contributions to the ground-state energy. The phase transition between the two ordered phases is first order for ferro- and anti-ferromagnetic Ising interactions, in agreement with previous works \cite{Lee04, Gammelmark11} but in contradiction to the mean-field calculation in Ref.~\cite{Zhang14}. We further extended the well-known Kramers-Wannier duality for the transverse-field Ising chain to the QTFIM resulting in the isospectral light-induced quantized Ising model in a tranverse field, which therefore also displays a first-order phase transition. In the future it would be interesting to extend our calculations to the case $\omega_0\neq 0$ so that only one $\mathbb{Z}_2$-symmetry remains and a second-order superradiant phase transition is known to be present. Another important aspect is a generalization of the light part, e.g.~Rabi lattice models with discrete gauge symmetry due to local photon operators are known to exhibit first- and second-order quantum phase transitions depending on the photon quantum dynamics \cite{nevado_2015}.

Experimentally, the QTFIM chain is expected to be realizable in a variety of quantum platforms~\cite{Lee04}. For example, a direct implementation within circuit QED has indeed already been given in \cite{Zhang14}. Other possible suitable candidates for experimental implementations could be within the field of opto-magnonics using YAG magnetic spheres in microwave cavities~\cite{Kusminskiy_16} or in ion traps, where the tailoring of collective motional modes can lead to a variety of spin models \cite{Cirac2005}. Alternatively, the manipulation of internal degrees of freedom with cavity quantum optical fields leads to an alternative platform where magnetism can be studied and observed as spinor ordering and texture formation in cold quantum gases~\cite{Mivehvar2017disorder,Mivehvar2019cavity,Kroeze2018spinor,Landini2018formation,Muniz2019exploring}.

In conclusion, the interplay between matter-matter and light-matter interactions can lead to a strong imprint of quantum cooperativity. Such opto-magnetic systems represent a promising playground for the discovery of yet unknown quantum phenomena and therefore hold the key for the emergence of quantum materials with enhanced capabilities.

\acknowledgments
 We acknowledge financial support from the Max Planck Society and from the German Federal Ministry of Education and Research, co-funded by the European Commission (project RouTe), project number 13N14839 within the research program "Photonik Forschung Deutschland" (C.~G.).

\onecolumngrid
\appendix

\section{The displaced Ising Hamiltonian}
The light-matter interaction in the Hamiltonian can be eliminated by a diagonalizing transformation that sees the field operators transformed as $ \hat{a} + \alpha \hat{S}_x = \hat{D}^\dagger \hat{a} \hat{D}$. The Rabi part consequently transforms to
	\begin{equation}
	\begin{aligned}
	\widetilde{\mathcal{H}}_\mathrm{Rabi} := \hat{D}\mathcal{H}_{\rm Rabi}\hat{D}^\dagger& = \omega_c \; \hat{D} \left(\hat{a}^\dagger + \alpha \hat{S}_x \right)\hat{D}^{\dagger}\hat{D}^{\phantom\dagger}\left(\hat{a}^{\phantom\dagger} + \alpha \hat{S}_x \right)\hat{D}^{\dagger} -\omega_c \alpha^2 \hat{S}_x^2  = \omega_c \left[a^\dagger a^{\phantom\dagger} - \alpha^2 \hat{S}_x^2\right].
	\end{aligned}
	\end{equation}
To transform $\mathcal{H}_{\rm Ising}$ we use the identity
	\begin{equation}
    \label{eq::smart_bch_formula}
	e^{\hat{A}}\hat{B}e^{-\hat{A}} = \hat{B} + [\hat{A},\hat{B}] + ... + \frac{1}{n!} \underbrace{[\hat{A},[\hat{A}, ...[\hat{A},\hat{B}]]...]}_{n\, \text{times}\, \hat{A}} + ...
	\end{equation}
Furthermore we introduce the abbreviations $\gamma\delta = \sum_i \sigma^\gamma_i\sigma^\delta_{i+1}$ and $\hat{r} = \alpha (\hat{a}^\dagger - \hat{a})$. It follows that
	\begin{equation}
	\begin{aligned}
\widetilde{\mathcal{H}}_\mathrm{Ising} := \hat{D} \mathcal{H}_{\rm Ising}\hat{D}^\dagger & = -J \left[ zz - \frac{\mathrm{i}}{2}(yz+zy)\sum_{k=0}^{\infty}\frac{1}{(2k+1)!}(2\hat{r})^{2k+1}+\frac{1}{2} (zz-yy) \sum_{k=1}^{\infty}\frac{1}{(2k)!}(2\hat{r})^{2k} \right ] \\
& = -J \left[ zz -\frac{\mathrm{i}}{2}(yz+zy)\sinh(2\hat{r}) + \frac{1}{2}(zz-yy)(\cosh(2\hat{r})-\mathds{1})\right ].
	\end{aligned}
	\label{TransformedH2}
	\end{equation}
The longitudinal field $\omega_0 S_z$ is transformed in the same way as the Ising Hamiltonian.
We use $S_z = \sum_i \sigma^z_i/2$ and do not write the constant prefactors.
Then, we obtain
\begin{equation}
  \begin{aligned}
    \hat{D}^{\phantom{\dagger}\hspace*{-1mm}}  \sum_i \sigma^z_i \hat{D}^\dagger &= \sum_i \exp(\alpha \hat{S}_x (\hat{a}^\dagger - \hat{a}^{\phantom{\dagger }})) \sigma^z_i
    \exp(-\alpha \hat{S}_x (\hat{a}^\dagger - \hat{a}^{\phantom{\dagger }})) \\
    &= \sum_i \sigma^z_i + (\alpha ({\hat{a}^\dagger - \hat{a}^{\phantom{\dagger }}}))[\sum_j \frac{\sigma^x_j }{2}, \sigma^z_i ] + \frac{1}{2!}(\alpha ({\hat{a}^\dagger - \hat{a}^{\phantom{\dagger }}}))^2 [\sum_k \frac{\sigma^x_k}{2}, [\sum_j \frac{\sigma^x_j}{2}, \sigma^z_i ]] + \ldots
  \end{aligned}
\end{equation}
where again \ref{eq::smart_bch_formula} is applied.
Since the photonic operators commute with all spin operators, these were written in front of the commutators.
The remaining spin commutators are only non-zero if the index $i$ of the outer sum is equal to all indices of the sums within the commutator.
Furthermore, one can find for the concatenated commutators
\begin{equation}
  [\sigma^x_i, \ldots [\sigma^x_i , [\sigma^x_i, \sigma^z_i ]]] =
  \left\{
  \begin{array}{cc}
    -\mathrm{i} \, 2^n \sigma^y_i  & n \text{ odd}\\
    2^n \sigma^z_i & n \text{ even}
  \end{array}\right.
\end{equation}
where $n$ is the number of Pauli-$x$ operators within the commutators.
The final result for the transformed is
\begin{equation}
  \begin{aligned}
    \hat{D}^{\phantom{\dagger}\hspace*{-1mm}}  \sum_i \sigma^z_i \hat{D}^\dagger &= \sum_{n=0} \mathrm{i}\, \frac{(\alpha (\hat{a}^\dagger - \hat{a}^{\phantom{\dagger}}))^{2n+1}}{(2n+1)!} \left(\sum_i \sigma^y_i \right)
    + \sum_{n=1} \frac{(\alpha( \hat{a}^\dagger - \hat{a}^{\phantom{\dagger}} ))^{2n}}{(2n)!} (\hat{a}^\dagger - \hat{a}^{\phantom{\dagger}})^{2n} \left( \sum_i \sigma^z_i \right)  \\
    &= \left( \sum_i \sigma^z_i \right) \cosh(\hat{r}) + \mathrm{i}\, \left(\sum_i \sigma^y_i\right) \sinh(\hat{r})
  \end{aligned}
\end{equation}

\section{Mapping of QTFIM onto TFIM in the strong coupling limit for the chain geometry}
We follow two approaches which both prove the mapping of the QTFIM onto the TFIM for the chain geometry under strong coupling, i.e.~high photon field amplitude conditions.
\subsection*{Approach 1: Neglecting non-extensive contributions}
\label{sec::qtfimtotfim}
To find the energies of the QTFIM in the limit where the Ising interaction is a perturbation, we will use the displaced basis and make use of Eq.~\eqref{TransformedH}. We will then notice that the perturbed energies can be exactly derived from the exact solution of the TFIM. For $J=0$ the two eigenstates of $\mathcal{H}_{\rm Rabi}$ are given by $\ket{\Psi}_{l} = \Ket{0} \otimes (\otimes_\nu\ket{\leftarrow}_\nu)$ and $\ket{\Psi}_{r} = \Ket{0} \otimes (\otimes_\nu\ket{\rightarrow}_\nu)$.  Treating $\mathcal{H}_{\rm Ising}$ as a perturbation the most important point to realize is that the terms
\begin{equation}
-\frac{\mathrm{i}}{2}(yz+zy)\sinh(2\hat{r})+ \frac{1}{2}(zz-yy)(\cosh(2\hat{r})-\mathds{1})
\end{equation}
with $\gamma\delta = \sum_i \sigma_\gamma^{(i)}\sigma_\delta^{(i+1)}$ and $\hat{r} = \frac{g}{\sqrt{N}\omega_c}(\hat{a}^\dagger - \hat{a}^{\phantom\dagger})$
do not give rise to extensive corrections of the ground-state energy for finite perturbation orders.
The reason for that is the factor $1/\sqrt{N}$ in $\hat{r}$.
Hence to obtain extensive perturbative corrections of the ground-state energy only the term $zz=\sum_i \sigma_z^{(i)}\sigma_z^{(i+1)}$ is relevant. We reformulate the problem as finding the ground-state energy per site $e_0=E_0/N$ of the Hamiltonian $\mathcal{H} = \widetilde{\mathcal{H}}_{\rm Rabi} - J \sum_{i=1}^N \sigma_{i}^z\sigma_{i+1}^z$ perturbatively in $J\omega_c/g^2$ for $N\rightarrow \infty$.
W.l.o.g. one can choose the eigenstate at order zero perturbation theory as $\ket{\Psi}_{r}$.
Since the magnetic and the photonic part of $\widetilde{\mathcal{H}}_{\rm Rabi}$ completely decouple we only write the magnetic part in the following derivation.

If $a$ spins of $\ket{\Psi}_{r}$ are flipped, the energy of $\widetilde{\mathcal{H}}_{\rm Rabi}$ is $-\frac{g^2}{4\omega_c N} (N-2a)^2 = N/4 (-\frac{g^2}{\omega_c}) + a \frac{g^2}{\omega_c} -a^2\frac{g^2}{\omega_cN}$. If $a$ is finite and $N\rightarrow \infty$ this is equal to $ N/4(-\frac{g^2}{\omega_c}) + a \frac{g^2}{\omega_c}$ and the spectrum of $\mathcal{H}_{\rm Rabi}$ is equidistant such that we can write
\begin{equation}
  \widetilde{\mathcal{H}}_{\rm Rabi} \approx -\frac{N g^2}{4 \omega_c} + \frac{g^2}{\omega_c} \sum_i \frac{\sigma_i^x}{2} + \frac{N g^2}{2 \omega_c} = \frac{N g^2}{4 \omega_c} + \frac{g^2}{\omega_c} \hat{S}_x  .
\end{equation}

Comparing with the transverse-field Ising chain
\begin{equation}
\mathcal{H}_{\rm TFIM} = -J \sum_i \sigma_i^{z}\sigma_{i+1}^z + h\hat{S}_x
\end{equation}
we then immediately see that for $h=g^2/\omega_c$ a perturbation in $J/h$ yields the same ground-state energy corrections in $\mathcal{H}_{\rm TFIM}$ as in the QTFIM. We conclude that given the perturbation series for the transverse-field Ising chain as $e_{0,\rm TFIM}(\lambda)$ with $\lambda = 2J/h$ as in Eq.~\eqref{eq::TFIM_e0_exact} the one for the QTFIM is
\begin{equation}
e_{0,\rm QTFIM}(J\omega_c/g^2) = e_{0,\rm TFIM}(2J\omega_c/g^2)+\frac{g^2}{4\omega_c}.
\end{equation}
In regions where the perturbative expansion converges this can be expressed with the exact formula for the ground-state energy of the Ising model in a transverse field \cite{Pfeuty} given by
\begin{equation}
e_{0,\rm QTFIM}(J\omega_c/g^2) = \frac{g^2}{4\omega_c}-\frac{g^2}{2\omega_c} \frac{1}{2\pi}\int_{0}^{2\pi}dk\sqrt{1+\lambda^2+2\lambda\cos(k)}.
\end{equation}
For $J = 0$, we obtain $e_{0,\rm QTFIM} = -g^2 / (4 \omega_c)$ which is consistent with the solution of the quantum Rabi Hamiltonian.

\subsection*{Approach 2: Mean-field approximation}
\label{sec::meanfieldapprox}

We now show that taking expectation values of the photonic part in the high-field phase leads to the same mapping as the previously described perturbative approach. To show this we use the original form of the Hamiltonian and assume that the photonic state of the system is a coherent state $\Ket{\tilde{\alpha}}$ with $\tilde{\alpha} = g\sqrt{N}/(2\omega_c)$. For the mean-field ansatz we write $\hat{a} = \Braket{\hat{a}} +\delta\hat{a}$ and $\hat{a}^\dagger = \Braket{\hat{a}} +\delta\hat{a}^\dagger$ where we assume that $\Braket{\hat{a}^\dagger} = \Braket{\hat{a}}$ as a consequence of coherent photonic states. This way we rewrite the QTFIM Hamiltonian as
\begin{equation}
\begin{aligned}
\mathcal{H}_{\rm QTFIM} & = \mathcal{H}_{\rm Ising} + \frac{g}{\sqrt{N}} \left(2\Braket{\hat{a}} +\delta\hat{a}\phantom \dagger + \delta\hat{a}^\dagger\right)\hat{S}_x + \omega_{\rm c}\;\left(\Braket{\hat{a}}^2 + \Braket{\hat{a}}(\delta\hat{a}\phantom \dagger + \delta\hat{a}^\dagger) + \delta\hat{a}^\dagger\delta\hat{a}\phantom \dagger\right).
\end{aligned}
\end{equation}
Because we are only interested in extensive contributions we discard all terms that are fluctuations of order $\sqrt{N}$. The expectation value $\Braket{\hat{a}} = \tilde{\alpha} = g\sqrt{N}/(2\omega_c)$. This yields
\begin{equation}
\mathcal{H}_{\rm QTFIM} \approx \mathcal{H}_{\rm Ising}  + \frac{g^2}{\omega_c} \hat{S}_x + \frac{g^2}{4\omega_c}N.
\end{equation}
This is the same result as obtained with the perturbative expansion.\\

\section{Perturbative expansion in the strong-coupling limit}
\label{sec::state_pert}

Let us compute in perturbation theory the corrections to the ground-state energy and the ground-state wave function for the case of weak Ising couplings $J \omega_c \ll g^2$.
For this we start with the analytically available solutions for the Rabi model and treat the Ising part as a perturbation.
For convenience, a photonic displacement operator is defined as follows:
\begin{equation}
  \hat{D}_\mathrm{ph}(x) = e^{x (\hat{a}^\dagger - \hat{a})}
\end{equation}
where $x \in \mathbb{R}$.
This operator acts only on the photonic part of the Hilbert space whereas the displacement $\hat{D}$ acts both on the spin and the photon part degrees of freedom.
Moreover, we introduce $\alpha = g/(\sqrt{N}\omega_c)$.

\subsection{Perturbation theory in the bare basis}

The eigenbasis of the Rabi Hamiltonian can be written as
\begin{equation}
  |\psi_{n,m, l}^{(0)}\rangle = \hat{D}^\dagger \left( |n\rangle \otimes |m, l\rangle \right) = \hat{D}_\mathrm{ph}^\dagger(m \alpha) |n\rangle \otimes |m, l\rangle
\end{equation}
where $|n\rangle$ are the eigenstates of $a^\dagger a$ and $S_x |m, l\rangle = m |m, l\rangle$ with an index to take into account that the spin states might be degenerate.
The corresponding energy levels are
\begin{equation}
  E_{n,m,l}^{(0)} = n\omega_c - \frac{g^2}{N \omega_c}m^2 = n\omega_c - \omega_c \alpha^2 m^2
\end{equation}
Apparently, the ground-state is degenerate twice.
However, for simplicity we do not consider this degeneracy since the error due to this assumption vanishes for large $N$.
Instead, the correction for the ground state $| \psi^{(0)}_+ \rangle := | \psi^{(0)}_{0,N/2,0} \rangle$ with positve $m = N/2$ is calculated
\begin{equation}
  \label{eq::pert_ansatz}
  \begin{aligned}
  |\psi^{(1)}_{+} \rangle := |\psi^{(1)}_{0,N/2,0} \rangle = \sum_{n = 0}^\infty\sum_{m = -N/2+1}^{N/2}\sum_l \frac{\langle m, l | D H_\mathrm{Ising} D^\dagger |0, N/2, 0\rangle}{E_{0, N/2, 0}^{(0)} - E_{n, m, l}^{(0)}} D^\dagger |n, m, l\rangle \\
  = -J \sum_{n = 0}^{\infty} \sum_{\substack{m, l\\m \ne -N/2}} \sum_{i = 1}^N \langle m, l | \sigma_i^z \sigma_{i+1}^z | N/2, 0 \rangle \frac{\langle n |\hat{D}_\mathrm{ph}(m \alpha) D^\dagger_\mathrm{ph}(\alpha N/2)|0\rangle}{-\omega_c \left(\frac{\alpha N}{2}\right)^2 - n \omega_c + \omega_c \alpha^2 m^2}\hat{D}_\mathrm{ph}^\dagger(\alpha m) |n\rangle \otimes |m, l\rangle
\end{aligned}
\end{equation}
where in the second line the values for the energies and the Ising operator were inserted and the magnetic and the photonic part of the scalar product were separated.

The sum above can be simplified by considering the action of the z-Pauli matrices on the state $|N/2, 0\rangle = |\rightarrow\rightarrow\ldots\rangle$.
We introduce the notation
\begin{equation}
  \label{eq::def_nu}
  | \nu \rangle := |\rightarrow\,\rightarrow\,\ldots\,\rightarrow\,\underset{\nu}{\leftarrow}\,\underset{\nu+1}{\leftarrow}\,\rightarrow\,\ldots\rangle = \sigma_\nu^z \sigma_{\nu +1}^z |\rightarrow\rightarrow\ldots\rangle.
\end{equation}
Therefore, the scalar product $\langle m, l | \sigma_i^z \sigma_{i+1}^z | N/2, 0 \rangle$ is only non-zero if $|m, l\rangle = |\nu\rangle$ for some $\nu = i$ and thus
\begin{equation}
  \label{eq::ml_simplification}
  \begin{aligned}
    \sum_i \sum_{m, l} \langle m, l | \sigma_i^z \sigma_{i+1}^z | N/2, 0 \rangle \frac{\langle n |\hat{D}_\mathrm{ph}(m \alpha) D^\dagger_\mathrm{ph}(\alpha N/2)|0\rangle}{-\omega_c \left(\frac{\alpha N}{2}\right)^2 - n \omega_c + \omega_c \alpha^2 m^2} \hat{D}_\mathrm{ph}^\dagger(\alpha m) |n\rangle\otimes|m, l\rangle = \\
    = \sum_i \sum_{m} \sum_{\nu} \delta_{m, N/2-2} \delta_{\nu, i} \frac{\langle n |\hat{D}_\mathrm{ph}(m \alpha) D^\dagger_\mathrm{ph}(\alpha N/2)|0\rangle}{-\omega_c \left(\frac{\alpha N}{2}\right)^2 - n \omega_c + \omega_c \alpha^2 m^2} \hat{D}_\mathrm{ph}^\dagger(\alpha m) |n\rangle|\nu\rangle = \\
    = \frac{\langle n |\hat{D}_\mathrm{ph}((N/2 - 2) \alpha) \hat{D}_\mathrm{ph}(-\alpha N/2)|0\rangle}{-\omega_c \left(\frac{\alpha N}{2}\right)^2 - n \omega_c + \omega_c \left(\alpha\frac{N-4}{2}\right)^2} \hat{D}_\mathrm{ph}^\dagger(\alpha (N/2-2)) |n\rangle \sum_\nu |\nu \rangle.
  \end{aligned}
  \end{equation}
This sum can be simplified by applying $\hat{D}(\alpha) \hat{D}(\beta) = \hat{D}(\alpha + \beta)$ for any pair of real numbers $\alpha, \beta$.
Furthermore, we will use the coherent state decomposition in terms of the number states \mbox{$\hat{D}(\alpha)|0\rangle = \exp \left(-|\alpha|^2/2\right) \sum_{n=0}^\infty \alpha^n/\sqrt{n!}|n\rangle$}.

Inserting \ref{eq::ml_simplification} into \ref{eq::pert_ansatz} gives
\begin{equation}
  \begin{aligned}
    |\psi^{(1)}_+ \rangle = -J \sum_{n = 0}^\infty\frac{\langle n |\hat{D}_\mathrm{ph}( - 2 \alpha)|0\rangle}{\omega_c \alpha^2 \left(4 - 2N \right) - n \omega_c} \hat{D}_\mathrm{ph}^\dagger(\alpha (N/2-2))|n\rangle \sum_\nu |\nu \rangle \\
    = -J \sum_n \frac{e^{-2|\alpha|^2} (-2\alpha)^n}{\sqrt{n!}}\frac{1}{\omega_c \alpha^2 \left(4 - 2N \right) - n \omega_c} \hat{D}_\mathrm{ph}^\dagger(\alpha (N/2-2))|n\rangle \sum_\nu |\nu \rangle
  \end{aligned}
\end{equation}
Note that the first order contribution vanishes for $N \rightarrow \infty$ since $\alpha \rightarrow 0$ and $\alpha^2 N = \text{const}$.

\subsection{Perturbation theory in the displaced basis}

An alternative perturbation theory calculation can be conducted on the displaced Hamiltonian, i.~e.~ the Hamiltonian
\begin{equation}
  \hat{D}^{\phantom{\dagger}\hspace*{-1mm}} \mathcal{H} \hat{D}^\dagger = \omega_c \hat{a}^\dagger \hat{a}^{\phantom{\dagger}} - \omega_c \alpha S_x^2 + \hat{D}^{\phantom{\dagger}\hspace*{-1mm}} \mathcal{H}_\mathrm{Ising} \hat{D}^\dagger
\end{equation}
with the Ising part as perturbation.
The unperturbed basis is $|n\rangle \otimes |m, l\rangle$ (same way as for the perturbation theory above).
Let us split the displaced Ising term into three parts:
\begin{equation}
 \hat{D}^{\phantom{\dagger}\hspace*{-1mm}} \mathcal{H}_\mathrm{Ising} \hat{D}^\dagger  = \underbrace{\frac{1}{2} \left(zz + yy\right)}_{\hat{V}_1} \underbrace{-\frac{\mathrm{i}}{2}(yz+zy)\sinh(\hat{r})}_{\hat{V}_2} + \underbrace{\frac{1}{2}(zz-yy) \cosh(\hat{r})}_{\hat{V}_3}
\end{equation}
The perturbation due to $\hat{V}_1$ is:
\begin{equation}
  - J \sum_{n, \nu} \frac{\langle n, \nu |\hat{V}_1 |0, N/2\rangle}{-\alpha^2 \left(\frac{N}{2}\right)^2 + -\alpha^2 \left(\frac{N-4}{2}\right)^2} |n\rangle \otimes |\nu\rangle = 0
\end{equation}
since
\begin{equation}
  \left(\sigma_\nu^z \sigma_{\nu + 1}^z + \sigma_\nu^y \sigma_{\nu + 1}^y \right)|\rightarrow\rightarrow\ldots\rightarrow\rangle = |\nu\rangle - |\nu\rangle
\end{equation}
where the state $|\nu\rangle$ is defined as in \ref{eq::def_nu}.
The perturbation due to $\hat{V}_2$ is:
\begin{equation}
   -J \sum_{n,\nu} \frac{\langle \nu | \hat{V}_2 | N/2 \rangle \langle n | \sinh(2\hat{r})| 0 \rangle}{\alpha^2 (4 - 2N) - n \omega_c}  |n\rangle \otimes |\nu\rangle = J \sum_n \frac{\langle n | \hat{D}_\mathrm{ph}(2\alpha) - \hat{D}_\mathrm{ph}^\dagger(2\alpha) | 0 \rangle}{2 \alpha^2 (4 - 2N) - n \omega_c} |n\rangle \otimes \sum_{\nu} | \nu \rangle
\end{equation}
since
\begin{equation}
  \left( \sigma_\nu^y \sigma_{\nu+1}^z + \sigma_\nu^z \sigma_{\nu + 1}^y\right) |\rightarrow\rightarrow\ldots\rightarrow\rangle = -2\mathrm{i} |\rightarrow\rightarrow\ldots\rightarrow\underset{\nu}{\leftarrow}\underset{\nu +1}{\leftarrow}\rightarrow \ldots \leftarrow\rangle = - 2\mathrm{i} |\nu\rangle
\end{equation}
and $\sinh{x} = \left( e^x - e^{-x} \right) / 2$.
An analogous way shows for $V_3$:
\begin{equation}
  -J \sum_{n,\nu} \frac{\langle \nu | V_3 | N/2 \rangle \langle n | \cosh(2\hat{r})| 0 \rangle}{\alpha^2 (4 - 2N) - n \omega_c}  |n\rangle \otimes |\nu\rangle = -J \sum_n \frac{\langle n | \hat{D}_\mathrm{ph}(2\alpha) + \hat{D}_\mathrm{ph}^\dagger(2\alpha) | 0 \rangle}{2 \alpha^2 (4 - 2N) - n \omega_c} |n\rangle \otimes \sum_{\nu} | \nu \rangle
\end{equation}
The sum of all three contributions is
\begin{equation}
  -J \sum_n \frac{\langle n | \hat{D}_\mathrm{ph}^\dagger(2\alpha) | 0 \rangle}{ \alpha^2 (4 - 2N) - n \omega_c} |n\rangle \otimes \sum_{\nu} | \nu \rangle = -J \sum_n \frac{e^{-2|\alpha|^2}(-2\alpha)^n}{\sqrt{n!}} \frac{1}{ \alpha^2 (4 - 2N) - n \omega_c} |n\rangle \otimes \sum_{\nu} | \nu \rangle
\end{equation}
which gives the same result for the first-order state correction as before if the inverse transformation is applied:
\begin{equation}
  \begin{aligned}
  | \psi^{(1)}_+ \rangle &= -J \sum_n \frac{e^{-2|\alpha|^2}(-2\alpha)^n}{\sqrt{n!}} \frac{1}{ \alpha^2 (4 - 2N) - n \omega_c} \hat{D}^\dagger \left(|n\rangle \otimes \sum_{\nu} | \nu \rangle\right) = \\
  &= -J\sum_n \frac{e^{-2|\alpha|^2} (-2\alpha)^n}{\sqrt{n!}}\frac{1}{\omega_c \alpha^2 \left(4 - 2N \right) - n \omega_c} \hat{D}_\mathrm{ph}^\dagger(\alpha (N/2-2))|n\rangle \sum_\nu |\nu \rangle
  \end{aligned}
\end{equation}

In order to calculate the photon distribution, we trace over the spin part of the total state $|\psi_+\rangle = | \psi^{(0)}_+ \rangle + | \psi^{(1)}_+ \rangle$ and obtain a density matrix
\begin{equation}
  \rho = \sum_{m,l} \langle m, l | \psi_+ \rangle \langle \psi_+ |  m, l \rangle = | \psi^{(0)}_{+, \mathrm{ph}} \rangle \langle \psi^{(0)}_{+, \mathrm{ph}} | + N | \psi^{(1)}_{+, \mathrm{ph}} \rangle \langle \psi^{(1)}_{+, \mathrm{ph}} |
\end{equation}
where the photonic states are defined as follows
\begin{equation}
  \begin{aligned}
    | \psi^{(0)}_{+, \mathrm{ph}} \rangle &:= e^{-\frac{|\alpha|^2 N^2}{8}}\sum_{n = 0}^\infty \frac{(-\alpha N)^n}{2^n \sqrt{n!}} | n \rangle \\
    | \psi^{(1)}_{+, \mathrm{ph}} \rangle &:= -J\sum_{n = 0}^\infty \frac{e^{-2|\alpha|^2} (-2\alpha)^n}{\sqrt{n!}}\frac{1}{\omega_c \alpha^2 \left(4 - 2N \right) - n \omega_c} \hat{D}_\mathrm{ph}^\dagger(\alpha (N/2-2))|n\rangle .
  \end{aligned}
\end{equation}
Then, the photon distribution $P(n)$ is
\begin{equation}
  \begin{aligned}
    P(n) = \Tr\left( | n \rangle \langle n |\rho \right) &= e^{-\frac{|\alpha|^2 N^2}{4}} \frac{(-\alpha N)^{2n}}{2^{2n} {n!}} +\\
    &+ J^2 \sum_{m=0}^\infty \langle m | n \rangle \langle n | \sum_{k, k^\prime = 0}^\infty \frac{e^{-2|\alpha|^2} (-2\alpha)^{k}}{\sqrt{k!} (\omega_c \alpha^2 \left(4 - 2N \right) - k \omega_c)} \frac{e^{-2|\alpha|^2} (-2\alpha)^{k^\prime}}{\sqrt{k^\prime!}(\omega_c \alpha^2 \left(4 - 2N \right) - k^\prime \omega_c)} \\
    &\cdot \hat{D}_\mathrm{ph}^\dagger(\alpha (N/2-2)) k\rangle \langle k^\prime | \hat{D}_\mathrm{ph}(\alpha (N/2-2)) m \rangle = \\
    &= e^{-\frac{|\alpha|^2 N^2}{4}} \frac{(-\alpha N)^{2n}}{2^{2n} {n!}} +\\
    &+ J^2 \sum_{k, k^\prime = 0}^\infty \frac{e^{-2|\alpha|^2} (-2\alpha)^{k}}{\sqrt{k!} (\omega_c \alpha^2 \left(4 - 2N \right) - k \omega_c)} \frac{e^{-2|\alpha|^2} (-2\alpha)^{k^\prime}}{\sqrt{k^\prime!}(\omega_c \alpha^2 \left(4 - 2N \right) - k^\prime \omega_c)} \\
    &\cdot \langle n | \hat{D}_\mathrm{ph}^\dagger(\alpha (N/2-2)) k\rangle \langle k^\prime | \hat{D}_\mathrm{ph}(\alpha (N/2-2))  n \rangle
  \end{aligned}
\end{equation}

\section{Conditional displacement and squeezing operators in the weak-coupling limit}
For $\alpha \ll 1$, i.e.~$g \ll \omega_c$ or large $N$, terms of order $\alpha^2$ or higher in the  displaced QTFIM Hamiltonian can be dropped (see Eqs.~\eqref{diagRabi} and \eqref{TransformedH} with $\omega_0 = 0$).
Then
\begin{equation}
  \begin{aligned}
  \hat{D}^{\phantom{\dagger }} \mathcal{H}_\mathrm{QTFIM} \hat{D}^\dagger &\approx \omega_c \left[\hat{a}^\dagger \hat{a}^{\phantom{\dagger}} - \alpha^2 \hat{S}_x^2 \right] - J \left( \sum_{\langle i, j \rangle} \sigma^z_i \sigma^z_j - \alpha\, \mathrm{i}\left(\hat{a}^\dagger - \hat{a}^{\phantom{\dagger }}\right) \left[\sum_{\langle i, j \rangle} \sigma^y_i \sigma^z_j + \sigma^z_i \sigma^y_j\right]\right) \\
  & = \omega_c \left[ \hat{a}^\dagger \hat{a}^{\phantom{\dagger}} - \alpha^2 \hat{S}_x^2 \right] + -J\left(zz - \alpha\, \mathrm{i}\left(\hat{a}^\dagger - \hat{a}^{\phantom{\dagger }}\right) (yz + zy)\right)
\end{aligned}
\end{equation}
with the same notation as used before.
For aesthetic resaons, one can change $\hat{a} \rightarrow \mathrm{i}\, \hat{a}$ without changing eigenvalues.
The key point however is that
\begin{equation}
  \begin{aligned}
    \left[\hat{S}_x^2, yz + zy\right] &= 0\\
    \left[zz, yz + zy\right] &= 0
  \end{aligned}
\end{equation}
and therefore, we can define a conditional displacement
\begin{equation}
  \hat{D}_\mathrm{yz} \propto \exp{(yz+zy) (\hat{a}^\dagger - \hat{a}^{\phantom{\dagger}})}
\end{equation}
which transforms the Hamiltonian into a diagonal form.
This principle might also work for higher orders, such as $\alpha^2$ where also terms describing two-photon generation are included.
However, another operator than a conditional displacement must be used in order to diagonalize a Hamiltonian with squared annihilation and creation operators.

\section{Numerical considerations: convergence}

\begin{figure}
	\centering
	\begin{minipage}{0.45\columnwidth}
		\centering
		\includegraphics[width=\textwidth]{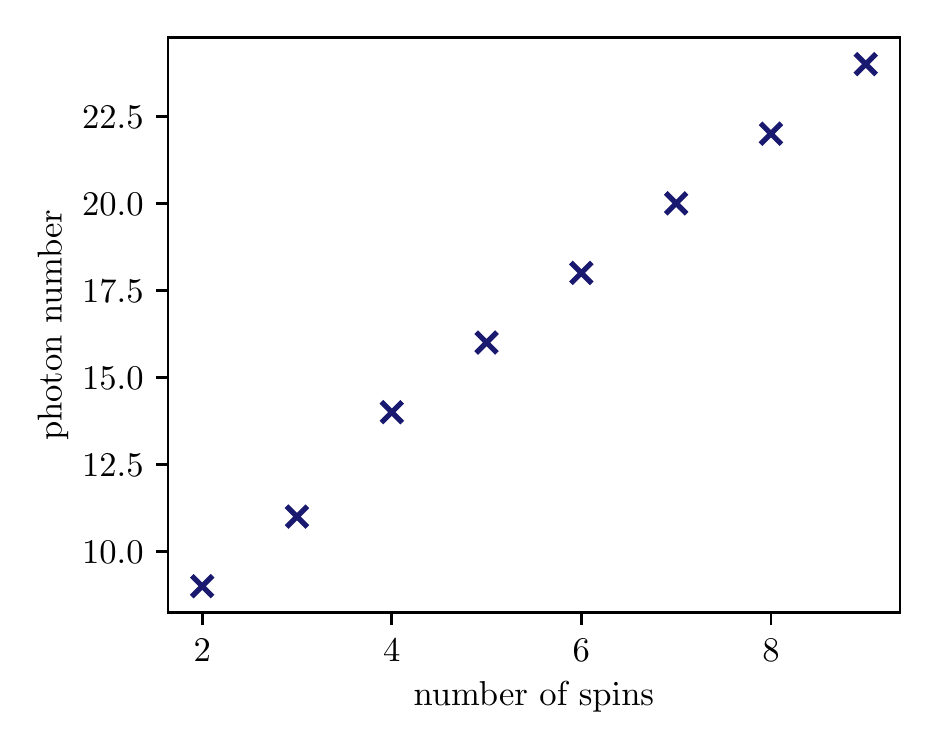}\\
		\textbf{a) fixed $J$ and $g$}
	\end{minipage}
\begin{minipage}{0.45\columnwidth}
		\centering
		\includegraphics[width=\textwidth]{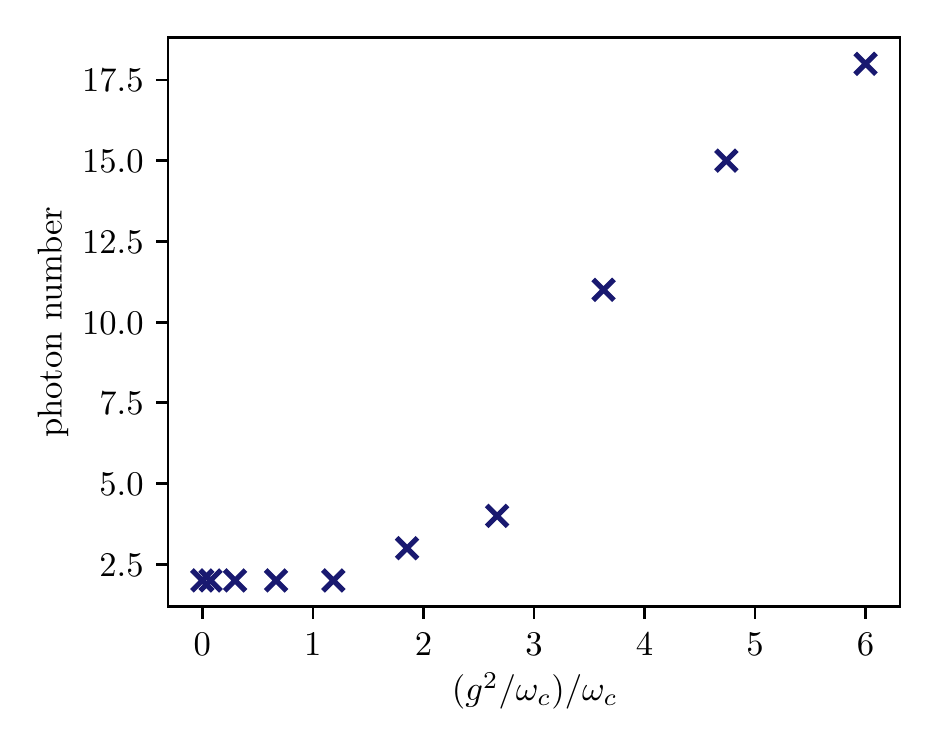}\\
		\textbf{b) fixed chain size}
	\end{minipage}
	\caption{a) Required cutoff of the photon number in order to reach an accuracy of $\delta=0.01\omega_c$ depending on the number of spins. b) Effect of the light-matter coupling constant $g$ on the required maximum photon number for a chain of 5 spins and $J/\omega_c = 0$.
	}
	\label{fig::convg}
\end{figure}

The analytical results are supported by numerical computations.
Since the photonic part of the Hilbert space is infinite, only the subspace including states with less than $n_\mathrm{max}$ photons is used for the matrix representation of the Hamiltonian.
Then, the ground state and the ground-state energy can be determined by finding the lowest eigenvalue of $L \times L$-matrices with $L \equiv 2^N \cdot n_\mathrm{max}$.
The quality of the approximation was ensured by measuring the maximal difference of the ground-state energy for $n_\mathrm{max} = n_\mathrm{test}$ and $n_\mathrm{max} = n_\mathrm{test}+1$ which should approach zero as $n_\mathrm{max}$ is increased.
Fig.~\ref{fig::convg} shows the computational effort for a fixed $\delta$.

Since for $\omega_0 = 0$ the ground state is two-fold degenerate (there is one state with magnetization $M_z = +M$ and another one with $M_z = -M$), the numerically obtained magnetiaztion $M_z$ can vary between $-M$ and $M$. Theoretically, the computed eigenvector can be an arbitray superposition of these two states with magnetization $\pm M$ which span the two-dimensional eigenspace corresponding to the extracted ground-state energy. In order to lift this degeneracy and only get the maximum magnetization $+M$, a small longitudinal field $\omega_0$ was added. $\omega_0$ was chosen small enough to still have a good approximation to the original QTFIM for $\omega_0=0$.

\end{document}